\documentclass[10pt]{article}
\usepackage{amsmath}
\usepackage{cite}
\usepackage{relsize}
\usepackage{amssymb}
\usepackage{graphics}
\usepackage{epsfig}
\usepackage{epstopdf}
\usepackage{verbatim}
\usepackage{color}
\usepackage{subcaption}
\usepackage{multirow}
\usepackage{textcomp}
\usepackage{float}
\usepackage{mathtools}
\usepackage[toc,page]{appendix}
\usepackage{enumitem}

\begin{document}
	\title{Magic neutrino mass model with broken $\mu-\tau$ symmetry and Leptogenesis}
	\author {Surender Verma\thanks{Electronic address: sverma@cuhimachal.ac.in, s\_7verma@yahoo.co.in} and Monal Kashav\thanks{Electronic address: monalkashav@gmail.com}}	
	\date{\textit{Department of Physics and Astronomical Science,\\Central University of Himachal Pradesh, Dharamshala 176215, INDIA.}}
	\maketitle
	\begin{abstract}
		We investigate baryogenesis via leptogenesis in $A_4$ flavor model within the paradigm of type-I and II seesaw mechanism resulting in magic neutrino mass matrix with broken $\mu-\tau$ symmetry in a minimal scenario with two right-handed neutrinos(2RHN). Additional $Z_3$ cyclic symmetry is employed to constrain the Yukawa structure of model. The type-II seesaw terms play crucial role in generating non-degenerate neutrino masses and non-zero $\theta_{13}$ and contribute in baryogenesis. In particular, after the spontaneous symmetry breaking, the Yukawa couplings $y_{\Delta_1}$ and $y_{\Delta_3}$ are responsible for the breaking of $\mu-\tau$ symmetry. The effective Majorana neutrino mass $|M_{ee}|$ is found to be well within the sensitivity reach of the $0\nu\beta\beta$ experiments, in particular, for inverted hierarchy. The model has imperative implication for inverted hierarchy, for example, the non-observation of this process at nEXO will rule out IH. The predicted baryon asymmetry is in good agreement with the observed baryon asymmetry for NH whereas IH is disallowed at 2.5$\sigma$ C.L.. 
		
	\end{abstract}
	\maketitle
	\section{Introduction}
	\par In the last two decades neutrino oscillation experiments have evinced that neutrino change identity as they travel from source to detector. This metamorphosis, in turn, requires that neutrinos are massive which, remarkably, is in contradiction with the prediction of the most celebrated theory in particle physics called ``Standard Model(SM)". Within SM, neutrinos are massless because (i) there are no right-handed(RH) neutrinos in the SM (ii) SM Lagrangian contains only renormalizable terms (iii) No Higgs triplet of $SU(2)_L$ in SM. In fact, Sudbury Neutrino Observatory(SNO)\cite{sno,sno2} and KamLAND\cite{kam,kam2} neutrino oscillation experiments have, decisively, demonstrated that neutrinos have tiny but non-zero mass and flavor eigenstates are different from the mass eigenstates. This observation has, since then, been augmented  by other experiments as well\cite{expt1,expt1a,expt2,expt2a,expt3}. Thus, the existence of non-zero neutrino mass is smoking gun signal of physics beyond SM(BSM). The mass-squared differences, mixing angles and Dirac-type $CP$ violation phase(s) can be probed in oscillation experiments, however, they are insensitive to the absolute neutrino mass scale. Consequent to the reasons described above, neutrino masses cannot be generated, the way quarks and charged lepton masses are generated within SM. Now, apart from identifying the underlying symmetry responsible for emerged picture of neutrino oscillation parameters(two mass-squared differences ($\Delta m_{12}^2$, $\Delta m_{23}^2$) and three mixing angles ($\theta_{12}, \theta_{23}, \theta_{13}$)) in neutrino oscillation experiments, the question is what mechanism is responsible for the origin of neutrino mass? In addition, there still remain experimental unknowns in the neutrino sector, some of which are:
	\begin{enumerate}
		\item Mass hierarchy-normal or inverted?
		\item Atmospheric mixing angle $\theta_{23}$- above or below maximality?
		\item Whether $CP$ symmetry is violated in neutrino oscillations?
		\item What is absolute scale of neutrino mass?
		\item Neutrino- Dirac or Majorana particle? etc.
	\end{enumerate}   
	The answers to first three questions are being investigated in the neutrino oscillation experiments like India-Based Neutrino Observatory(INO)\cite{ino}, Deep Underground Neutrino Experiment(DUNE)\cite{dune}, T2HK\cite{t2hk} and NO$\nu$A\cite{nova} etc. whereas direct search neutrino-mass-experiments\cite{katrin1,katrin2} and lepton number violating processes like 0$\nu\beta\beta$ decay will probe answers to the remaining two questions. These experiments will, further, help in ameliorating the lepton mixing paradigm. Thanks to the arduous experimental efforts that the dominant structure of the neutrino mixing matrix has been revealed and mass-squared differences, mixing angles are known to an unprecedented accuracy. The emerged picture of neutrino masses and mixing is dissimilar to the one in quark sector. For example, the quark mixing angles are small as compared to in neutrino sector(except $\theta_{13}$). In a more general framework like Grand Unified Theories(GUTs) in which quarks and leptons belong to the same representation of the gauge group and Yukawa couplings are related, it is a formidable task to understand such diasporical hierarchies in fermion masses and mixing.

	On the theoretical front, non-zero neutrino mass and the emerged spectrum of neutrino masses and mixing angles have posed outstanding fundamental questions before the theorists. In fact, there exist alternative theoretical mass models to explain neutrino parameters, for a review see Refs.\cite{mm1,mm2,mm3}. The reason for the possibility of variety of models is the unresolved experimental ambiguities, discussed in the last section, that still exist. We can broadly classify various model building frameworks as (i) theoretical models based on flavor symmetries to explain observed mixing parameters (ii) GUTs, as an attempt to unify quarks and leptons family structures (iii) models based on extra dimensions etc.. The physics of underlying flavor structure is contained in the Yukawa sector of the theory. The Yukawa couplings are, in general, complex and seeds the possible leptonic $CP$ violation. To explain the observed spectrum of neutrino masses and mixing, Yukawa sector needs to be extended through introduction of additional fermionic and scalar fields. One way to accomplish it, is the imposition of non-abelian discrete flavor symmetry group at some high energy scale and subsequent spontaneous breaking of the flavor symmetry in charged lepton and neutrino sector at low energy scale. In this approach, appropriate charge assignments to the fermionic and scalar field(s) under the flavor group alongwith vacuum alignment of the Higgs or Higgs like field(s) constrain the Yukawa structure of the flavor theory. In general, the proliferation of the field content leads to deterioration of the predictability of the model. Also, we may have to apply additional symmetry(ies) to avoid the unwanted couplings in the Lagrangian of the theory. So, we need to have a minimal extension of the fermionic and scalar sector to reduce free parameters in the theory.  
	
	The observation of large mixing is another unresolved puzzle in the neutrino sector. There exist myriads of mixing paradigms to explain observed mixing spectrum such as tri-bimaximal mixing(TBM), bi-maximal mixing, golden ratio and tri-maximal mixing, to name a few\cite{tbm1,tbm2,bm1,bm2,gr1,gr2,tm1,tm2}. TBM mixing predicts maximal atmospheric mixing($\theta_{23}=45^o$) and vanishing reactor angle($\theta_{13}=0$) and as such requires corrections to generate non-zero value of $\theta_{13}$ consistent with experimental observations at T2K\cite{t2k}, Daya bay\cite{db}, Double Chooz\cite{dc} and RENO\cite{reno}. The neutrino mass matrix resulting from TBM is, in general, ``magic" and ``$\mu-\tau$" symmetric which means that sum of elements of each row/column remains same and mass matrix is invariant under exchange of $\mu-\tau$ indices, respectively. 
	
	\par In the present work, we propose a neutrino mass model for the amendment of TBM ansatz such that $\mu-\tau$ symmetry is broken i.e. $\theta_{13}$ is non-zero, however, the mass matrix still has magic symmetry. In an effective theory layout, dimension-five operator instinctively generates small neutrino mass due to suppression by the cutoff scale($\Lambda$). Furthermore, the large mixing angles in leptonic sector hints toward additional $CP$ violation sources apart from the quark sector. The leptonic $CP$ violation sources may generate the observed baryon asymmetry of the universe(BAU) and have triggered arduous studies based on flavor models which, simultaneously, account for correct neutrino phenomenology as well. The matter-antimatter asymmetry is measured as baryon to photon ratio which through cosmological findings is\cite{bpr}
	$$|\eta_B|=(6.12\pm0.04) \times 10^{-10}.$$
	Baryon symmetric universe must satisfy the Sakharov's three necessary conditions
	\begin{itemize}
		\item baryon number violation
		\item $C$ and $CP$ violation
		\item out-of-equilibrium decay 
	\end{itemize}
	essential to generate observed baryon asymmetry\cite{sakharov}. The BSM scenarios with one or more right-handed neutrino, scalar Higgs triplet and/or scalar fermion triplet must account for observed neutrino phenomenology. Also, their decay contributes to $CP$ asymmetry and hence matter-antimatter asymmetry. Additional  BSM fields generates lepton asymmetry which is converted to baryon asymmetry through sphaleron transitions. Admittedly, the next ambitious goal is to understand, coetaneously, the mechanism of neutrino mass generation, large leptonic mixing and observed baryon asymmetry of the Universe(BAU). 
	\par Motivated by this, we present an attractive model based on $A_ 4$ flavor symmetry for generation of both neutrino masses and  leptogenesis in the framework of type-I and II seesaw mechanism. In this model, type-II seesaw plays a vital role in breaking $\mu-\tau$ symmetry and obtaining non-degenerate neutrino masses. The additional flavon fields introduced in the model drives appropriate breaking of the $A_4$ symmetry. We have, also, studied the leptogenesis in the type-I and II seesaw scenario using approximate solutions to the Boltzmann equations. 
	
	In Sec. II  we have discussed the $A_{4}\times Z_{3}$ model where $Z_3$ controls the Yukawa structure of the theory. In Sec. III and IV, we have presented the leptogenesis framework and constraining equations used in the numerical analysis, respectively. Finally, we present major conclusions of the work in Sec. V.

	\section{The $A_4$ Model}
	\begin{table}[t]
		\centering
		\begin{tabular}{cccccccccccccc}
			Symmetry & $\bar{D}_{iL}$ & $e_{R}$ & $\mu_{R}$ & $\tau_{R}$ & $N_1$ & $N_2$ &H &  $ \phi_{l}$  & $\phi_{\nu}$  &  $\chi_{1}$ &  $\chi_{2}$&   $\Delta$ \\ \hline
			$SU(2)_L$    &     2      &    1      &   1     &    1       & 1 & 1 &   2        &    1   & 1 & 1& 1     &3     \\ \hline
			$A_{4}$ &      3     &      1    &     1$'$      &     1$''$       & 1 & 1$''$  &    1       &    3   &     3   &      1& 1$''$  &  1$''$\\ \hline
			$Z_{3}$ & $\omega^{2}$ & 1 & 1 & 1 & $\omega$ & $\omega$ & 1& $\omega$ &1 & $\omega$ &  1 & $\omega^{2}$ \\ 
		\end{tabular}
		\caption{\label{tab1} Field content of the model and charge assignments under $SU(2)_L$, $A_{4}$ and $Z_3$.} 
	\end{table}
	$A_{4}$ is a non-Abelian discrete group of even permutations of four objects having geometrical resemblance with tetrahedron.

	A$_4$ has twelve elements and four irreducible representations $\textbf{1}, \textbf{1}^{'},\textbf{1}^{''}$ and $\textbf{3}$. In the singlet representation, the generators S and T are:
\begin{eqnarray}
\nonumber
\textbf{1}:&& S=1, T=1,\\ \nonumber
\textbf{1}^{'}:&& S=1, T=\omega^{2},\\ \nonumber
\textbf{1}^{''}:&& S=1, T=\omega,\\
\end{eqnarray}
where $\omega=e^{2\pi i/3}$. For triplet vector space, we have chosen T-diagonal basis in which the generators S and T are given by

  \begin{equation}
  \text{S}=\frac{1}{3}\begin{pmatrix}
-1 & 2 &  2\\
2 & -1 &  2\\
2 & 2 &  -1\\
 \end{pmatrix}\hspace{1cm}\text{and}\hspace{1cm}\text{T}=\begin{pmatrix}
1 & 0 &  0\\
0 & \omega^2 &  0\\
0 & 0 &  \omega\\
 \end{pmatrix}.
   \end{equation}
The generators satisfy $\text{S}^{2}=\text{T}^{3}=\text{(ST)}^{3}=1$. The representations multiplication rules are: $\textbf{1}\otimes$ other$=$other, $\textbf{1}^{'}\otimes\textbf{1}^{'}=\textbf{1}^{''}$, $\textbf{1}^{'}\otimes\textbf{1}^{''}=\textbf{1}$, $\textbf{1}^{'}\otimes\textbf{3}=\textbf{3}$, $\textbf{1}^{''}\otimes\textbf{1}^{''}=\textbf{1}^{'}$, $\textbf{1}^{''}\otimes\textbf{3}=\textbf{3}$, $\textbf{3}\otimes\textbf{3}=\textbf{1}\oplus\textbf{1}^{'}\oplus\textbf{1}^{''}\oplus\textbf{3}_s\oplus\textbf{3}_a$. In this basis, the Clebsch-Gordon decomposition of two triplets $a=(a_1,a_2,a_3)$ and $b=(b_1,b_2,b_3)$ is given as

\begin{eqnarray}
\centering
\nonumber
&&(a\otimes b)_{\textbf{1}}=a_{1}b_{1}+a_{2}b_{3}+a_{3}b_{2},\\ \nonumber
&&(a\otimes b)_{\textbf{1}^{'}}=a_1b_2+a_2b_1+a_3b_3,\\ \nonumber
&&(a\otimes b)_{\textbf{1}^{''}}=a_1b_3+a_2b_2+a_3b_1,\\ \nonumber
&&(a\otimes b)_{\textbf{3}_{s}}=\frac{1}{3}(2a_1b_1-a_2b_3-a_3b_2,2a_3b_3-a_1b_2-a_2b_1,2a_2b_2-a_1b_3-a_3b_1), \\ \nonumber
&&(a\otimes b)_{\textbf{3}_{a}}=\frac{1}{2}(a_2b_3-a_3b_2,a_1b_2-a_2b_1,a_1b_3-a_3b_1).
\end{eqnarray}
	 Here we present $A_{4}\otimes Z_3$ model within type-I and II seesaw framework. In this model, we have employed one $SU(2)_{L}$ 
	Higgs doublet H, $SU(2)_L$ singlet flavon fields $\phi_{l},\phi_{\nu},\chi_{1},\chi_2$ and one $SU(2)_{L}$ triplet Higgs fields($\Delta$). The transformation properties of different fields under $SU(2)_{L}$, $A_4$ and $Z_3$ are given in Table \ref{tab1}. The charge assignments under $SU(2)_L$, $A_4$ and $Z_3$ lead to the following invariant Yukawa Lagrangian

	\begin{eqnarray}
	\nonumber
	- \mathcal{L}= && \frac{y_{e}}{\Lambda}\bar{D}_{iL} H \phi_{l} e_{R} + \frac{y_{\mu}}{\Lambda}\bar{D}_{iL} H \phi_{l} \mu_{R} + \frac{y_{\tau}}{\Lambda}\bar{D}_{iL} H \phi_{l} \tau_{R} \\
	\nonumber
	&& + \frac{y_{\nu_{1}}}{\Lambda}\bar{D}_{iL} \tilde{H} \phi_{\nu} N_{1} +\frac{y_{\nu_{2}}}{\Lambda}\bar{D}_{iL} \tilde{H} \phi_{\nu} N_{2}\\
	\nonumber
	&& - h_{\chi_ 1} N^{T}_{1} C^{-1} N_{1}\chi_1 -  \frac{h_{\chi_ 2}}{\Lambda} \chi_{1}\chi_{2} N^{T}_{2} C^{-1} N_{2}\\
	\nonumber
	&&  -y_{\Delta_{1}}\bar{D}_{iL}  \Delta {D}_{iL}^{c}-\frac{y_{\Delta_{2}}}{\Lambda}\bar{D}_{iL}  \Delta {D}_{iL}^{c}\phi_{\nu} -\frac{y_{\Delta_{3}}}{\Lambda}\bar{D}_{iL}  \Delta {D}_{iL}^{c}\chi_{2}+ H.c.,\\  \label{lag}
	\end{eqnarray}
	where $\tilde{H}$=i$\tau_2 H^*$($H$ is SM Higgs doublet) and $y_{i}$ ($i= e$,$\mu, \tau, \nu_{1}, \nu_{2}, \chi _1, \chi_2, \Delta_{1} ,\Delta_{2},\Delta_{3}$) are Yukawa coupling constants. We have considered the vacuum expectation values($vev$) for flavon fields $\phi_l$ and $\phi_\nu$ as $(v_{l},0,0)$ and 
	$(v_{\nu},v_{\nu},v_{\nu})$, respectively  and $vev$ for $SU(2)_{L}$  Higgs doublet to be $v_{H}$.  In T-diagonal basis, the Lagrangian in Eqn.(\ref{lag}) results in diagonal charged lepton mass matrix

	\begin{equation} \label{lepton}
	m_{l}=\frac{v_{l}v_{H}}{\Lambda}
	{\begin{pmatrix}
		{y_{e}}& 0&0 \\
		0  & {y_{\mu}}& 0 \\
		0& 0 &{y_{e}}  
		\end{pmatrix}},
	\end{equation}
	and Dirac neutrino mass matrix
	
	\begin{equation} \label{dirac}
	m_{D}=\frac{v_{\nu}v_{H}}{\Lambda}
	{\begin{pmatrix}
		y_{\nu_{1} }& y_{\nu_{2} }\\
		y_{\nu_{1} } & y_{\nu_{2} } \\
		y_{\nu_{1} }&y_{\nu_{2} }  
		\end{pmatrix}}.
	\end{equation}
	Also, the right-handed neutrino mass matrix is given by
	\begin{equation} \label{right}
	m_{R}=
	{\begin{pmatrix}
	 y_{\chi_{1}} v_{\chi_{1}}& 0\\
		0 & \frac{y_{\chi_{2}}}{\Lambda} v_{\chi_{1}}  v_{\chi_{2}}
		\end{pmatrix}},
	\end{equation}
	where $v_{\chi_{1}}$ and $v_{\chi_{2}}$ are the vacuum expectation values of the flavon fields $\chi_1$ and $\chi_2$, respectively.
	Using type-I seesaw relation, the effective neutrino mass matrix is given by
	\begin{equation} \label{type1}
	m_{\nu_{1}} =-m_{D} m_{R}^{-1} m_{D}^{T} =  
	{\begin{pmatrix}
		c& c&c\\
		c& c&c \\
		c& c&c  
		\end{pmatrix}},
	\end{equation}
	where $c=\frac{a^{2}}{M_{1}}+\frac{b^{2}}{M_{2}}$ with
	\begin{equation} \label{redefine}
	a=y_{\nu_{1} } \frac{v_{\nu}v_{H}}{\Lambda}, b=y_{\nu_{2} } \frac{v_{\nu}v_{H}}{\Lambda}, M_{1}= y_{\chi_{1}} v_{\chi_{1}}, M_{2}=\frac{y_{\chi_{2}}}{\Lambda} v_{\chi_{1}}v_{\chi_{2}}.
	\end{equation}
	$M_{1}$ and $M_{2}$ are the masses of right-handed neutrinos $ N_{1}$ and $N_{2}$, respectively.
	The mass matrix in Eqn.(\ref{type1}) results in degenerate light neutrino masses and $\theta_{13}=0$. In order to have correct low energy phenomenology we consider type-II seesaw scenario in conjunction with type-I. The conjoining of type-II seesaw terms have interesting implications on neutrino mass and leptogenesis. Assuming $vev$ for scalar triplet field $\Delta$ to be $v_{\Delta}$ the type-II contribution to effective neutrino mass matrix is given by
	\begin{equation} \label{type2}
	m_{\nu_{2}} = 
	{\begin{pmatrix}
		2p& d-p&f-p\\
		d-p&f+2p&-p \\
		f-p& -p&2p+d 
		\end{pmatrix}}
	\end{equation}
	where, $d=y_{\Delta_{1}}v_{\Delta}$, $ p=  \frac{y_{\Delta_{2}}}{3\Lambda}v_{\Delta} v_{\nu}$ and $f=\frac{y_{\Delta_{3}}}{\Lambda}v_{\Delta}v_{\chi_{2}} $. The charged lepton mass matrix in Eqn.(\ref{lepton}) is diagonal therefore effective neutrino mass matrix is given by
	 $$m_{\nu}=m_{\nu_{1}}+m_{\nu_{2}}.$$ 
	Using  Eqns.(\ref{type1}) and  (\ref{type2}) we get
	\begin{equation} \label{total}
	m_{\nu} = 
	{\begin{pmatrix}
		c+2p& c+d-p&c-p+f\\
		c+d-p&c+2p+f&c-p \\
		c-p+f& c-p&c+2p+d 
		\end{pmatrix}}.
	\end{equation}
	
	With $f=0$, Eqn.(10) can be written as
	
	\begin{equation} 
	m_{\nu} = 
	{\begin{pmatrix}
		c+2p& c+d-p&c-p\\
		c+d-p&c+2p&c-p \\
		c-p& c-p&c+2p+d 
		\end{pmatrix}}
	= \underbrace{{\begin{pmatrix}
			c+2p& c-p&c-p\\
			c-p&c+2p&c-p \\
			c-p& c-p&c+2p 
			\end{pmatrix}}
	}_{m_{\nu}^a}+
	\underbrace{  {\begin{pmatrix}
			0& d&0\\
			d&0&0 \\
			0& 0&d 
			\end{pmatrix}},
	}_{m_{\nu}^b}
	\end{equation}
	where $m_{\nu}^a$ is magic and $\mu-\tau$ symmetric and $m_\nu^b$ breaks the $\mu-\tau$ symmetry. Interestingly, the breaking pattern obtained here viz.,
	\begin{equation}
	\begin{pmatrix}
	0& d&0\\
	d&0& 0\\
	0& 0&d 
	\end{pmatrix}, \nonumber
	\end{equation}
	is proposed in Ref.\cite{magic}. Here, we have shown the symmetry origin of such breaking pattern for $\mu-\tau$ symmetry while retaining magic symmetry. However, in this case $m_\nu$(Eqn.(11)) contains two equalities viz. $(m_{\nu})_{11}=(m_{\nu})_{22}$ and $(m_{\nu})_{13}=(m_{\nu})_{23}$ which is disallowed by current data on neutrino mass and mixings\cite{hybrid}. So, in the following we have considered $f\ne 0$.
	
The neutrino mass matrix in Eqn.(\ref{total}) obey magic symmetry and is $\mu-\tau$ asymmetric. The dimension-four and dimension-five terms in Eqn.(\ref{lag}) involving the scalar triplet $\Delta$ provide non-degenerate light neutrino masses. After the spontaneous symmetry breaking the Yukawa couplings $y_{\Delta_1}$ and $y_{\Delta_3}$(contained in $d$ and $f$, respectively) are responsible for the breaking of $\mu-\tau$ symmetry. To see this explicitly we can write Eqn.(\ref{total}) as
	\begin{equation} 
	m_{\nu} =
	\underbrace{{\begin{pmatrix}
			c+2p& c-p&c-p\\
			c-p&c+2p&c-p \\
			c-p& c-p&c+2p 
			\end{pmatrix}}
	}_{m_{\nu}^c}+
	\underbrace{  {\begin{pmatrix}
			0& d&f\\
			d&f&0 \\
			f& 0&d 
			\end{pmatrix}}.
	}_{m_{\nu}^d}
	\end{equation}
	$m_{\nu}^c$ is magic and $\mu-\tau$ symmetric. $m_\nu^d$ breaks the $\mu-\tau$ symmetry and makes $(m_{\nu})_{11}\ne(m_{\nu})_{22}$ and $(m_{\nu})_{13}\ne(m_{\nu})_{23}$ to have an accordant neutrino phenomenology.

	In the next section we have discussed the predictions of the model given in Eqn.(10) for beyond neutrino sector observables like effective Majorana neutrino mass $|M_{ee}|$ and leptogenesis and, also, investigated possible connection between breaking parameters ($d$, $f$) and the $CP$ asymmetry.  
	
	\section{Leptogenesis}
\begin{figure} [t] 
		\includegraphics[scale=0.25]{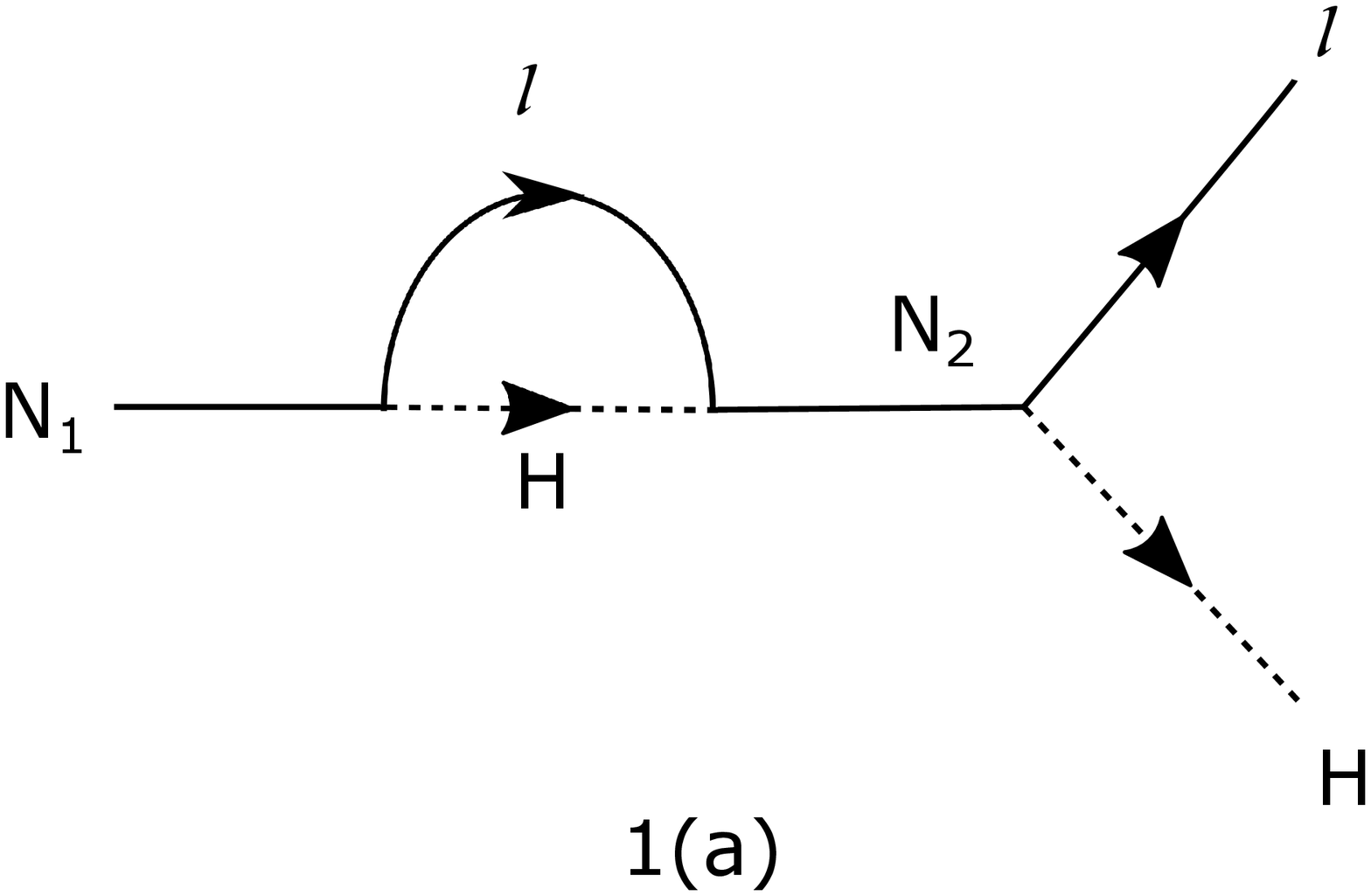}
		\includegraphics[scale=0.25]{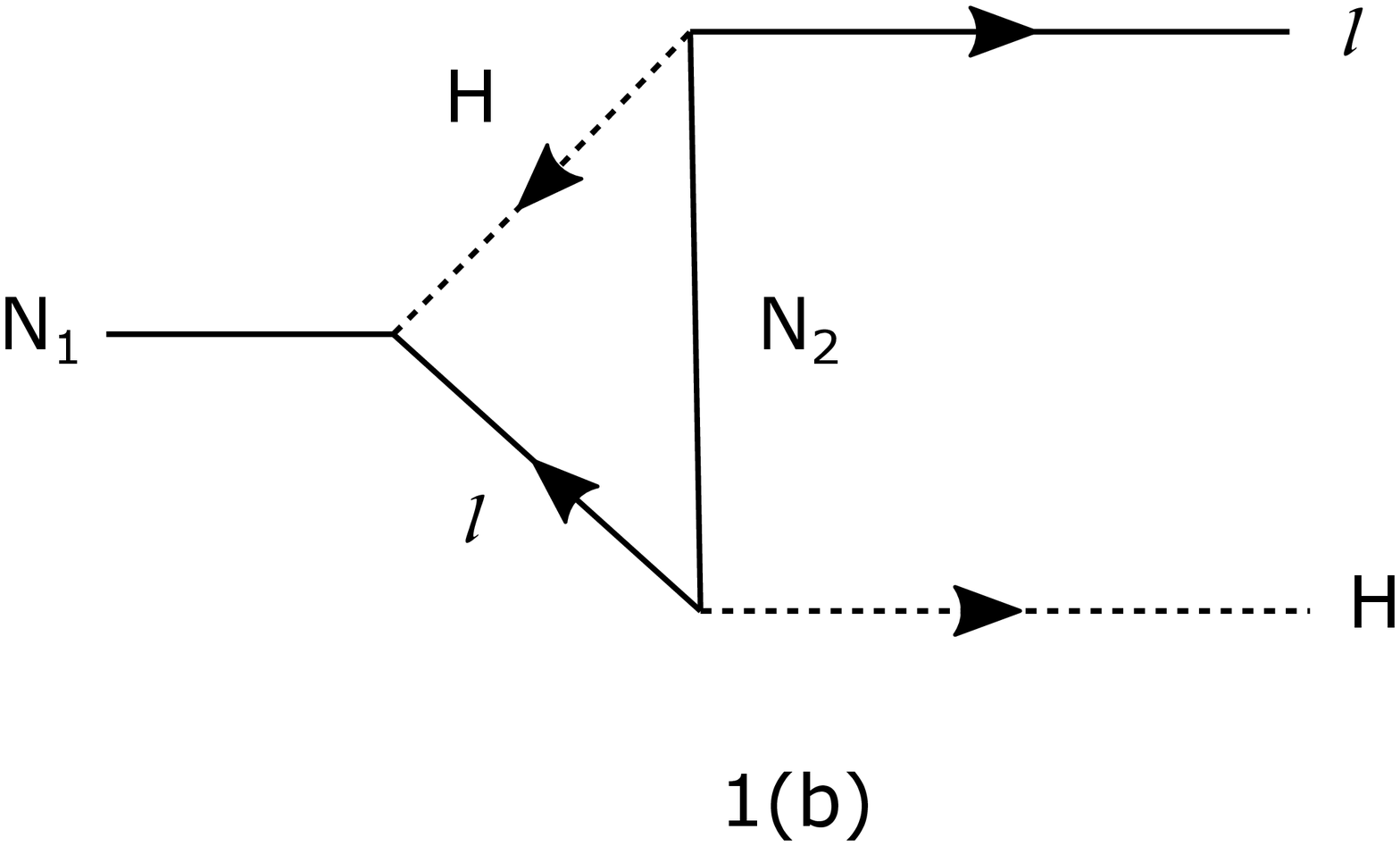}
		\includegraphics[scale=0.25]{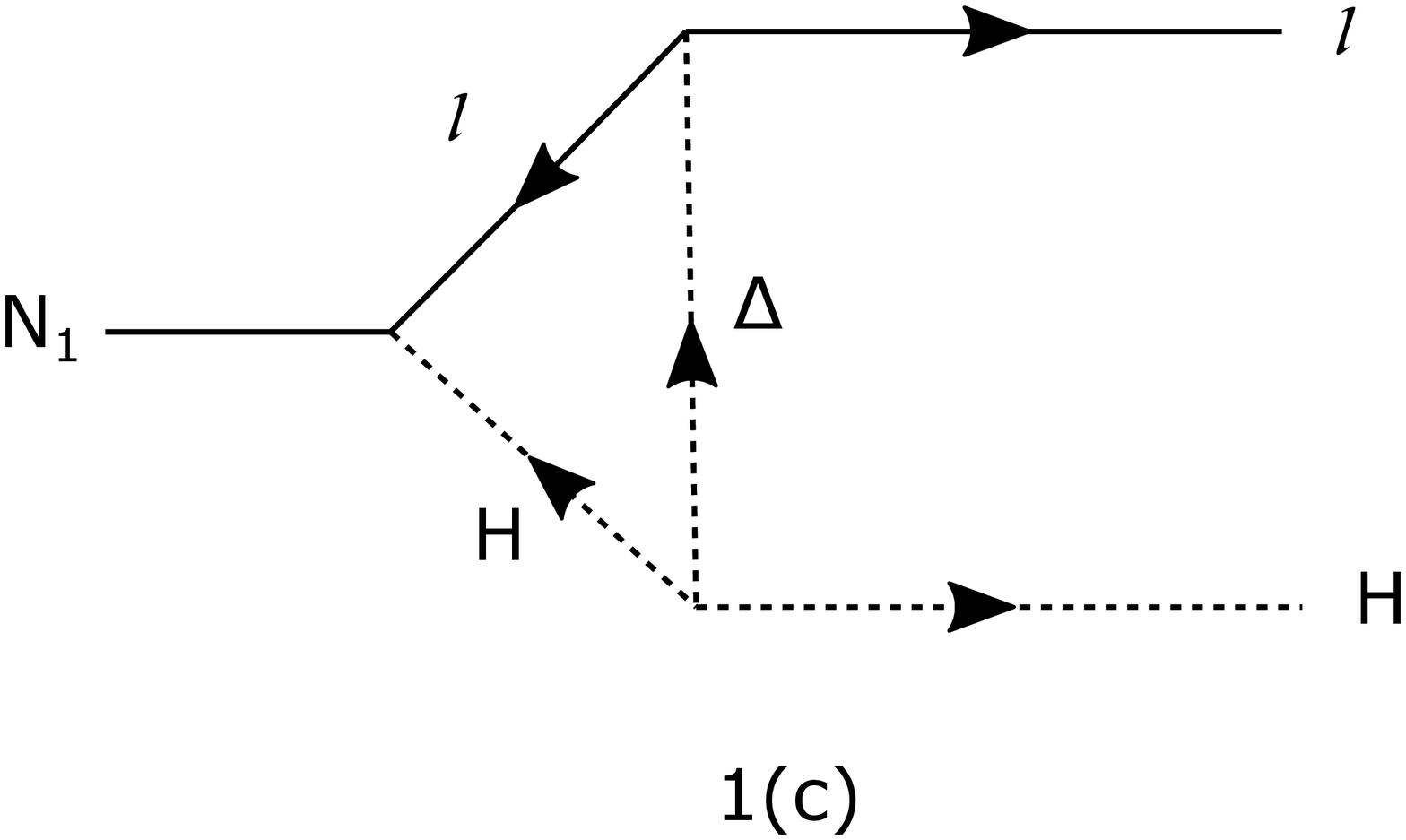}
		\caption{~\label{fig1} One loop-level diagrams contributing to $CP$ asymmetry.}
	\end{figure}

	The basic ingredients required for the leptogenesis are (i) $CP$ asymmetry induced by the decay of right-handed neutrino and scalar triplet $\Delta$ (ii) efficiency factor ($K_{eff}$) allowing to calculate the lepton asymmetry induced by the resulting $CP$ asymmetry, and (iii) the conversion of lepton asymmetry into baryon asymmetry through the $B-L$ conserving sphaleron processes in the SM. In addition to the right-handed neutrino decay, the scalar triplet $\Delta$ interactions, also, contribute to the net $CP$ asymmetry produced in the early universe. However, for hierarchical mass spectrum of right-handed neutrinos i.e. $M_1<<M_2$ and $M_1<<M_{\Delta}$, the decay of lightest right-handed neutrino($N_{1}$) is dominating.  It follows that while  $N_{2}$ and $\Delta$ decays, the interactions involving $N_{1}$ were in thermal equilibrium. The baryon asymmetry contribution due to decay of $N_{2}$ and $\Delta$ is washed out  before $N_{1}$ decays. The different decay modes at one loop-level correction are shown in Fig.(\ref{fig1}). The contribution due to out-of-equilibrium decay of lightest right-handed neutrino($N_{1}$) will survive. $CP$ asymmetry due to interference of tree level and one loop level decay amplitudes involving right-handed neutrino as shown in Fig.1(a) and Fig.1(b) can be written as\cite{eps1,eps3}
	\begin{equation}\label{epsilonN}
	\epsilon_{N}=\frac{3 M_1}{8\pi v_{H}^{2}} \frac{Im[m_{D}^{\dagger}m_{\nu_1}m_{D}^{*}]_{11}}{(m_{D}^{\dagger}m_{D})_{11}},
	\end{equation}
	and contribution from scalar triplet Higgs via Fig.1(c) is
	\begin{equation}\label{epsilonD}
	\epsilon_{\Delta}=\frac{3 M_1}{8\pi v_{H}^{2}} \frac{Im[m_{D}^{\dagger}m_{\nu_2}m_{D}^{*}]_{11}}{(m_{D}^{\dagger}m_{D})_{11}}.
	\end{equation}
	For out-of-equilibrium decay, the decay rate of $N_{1}$ i.e. $\Gamma_{1}$  must be smaller than expansion rate of the universe at temperature $T=M_{1}$ which can be expressed as 
	\begin{equation}\label{planck}
	\frac{(m_{D}^{\dagger}m_{D})_{11}}{M_1} <\sqrt\frac{64g^{*}\pi^5}{45}.\frac{v_{H}^{2}}{M_{pl}}\approx 1.08\times10^{-3}eV,
	\end{equation}  
	where $M_{pl}$ is Planck mass and $g*$ is total number of relativistic degree of freedom\cite{buch}.
	The baryon asymmetry produced via the sphaleron processes using full set of Boltzmann equations can be, approximately, written as
	\begin{equation}\label{eta}
	|\eta_B|\approx 0.96\times 10^{-2} \epsilon K_{eff},
	\end{equation}
	where $\epsilon=\epsilon_{N}+\epsilon_{\Delta}$ and  $K_{eff}$ is known as efficiency factor given by
	\begin{equation} \label{efficiency}
	K_{eff}=2\times10^{-2}\left(\frac{M_1\times0.01eV}{(m_{D}^{\dagger}m_{D})_{11}}\right)^{1.1},
	\end{equation}
	and measures the washout effects on lepton number asymmetry.
	
	Using Eqns.(\ref{dirac}), (\ref{type1})  and (\ref{epsilonN}) we obtain 
	\begin{equation}\label{epsilonN1}
	\epsilon_{N}=\frac{ 9M_1}{8\pi v_{H}^{2}}\frac{|b^{2}|}{|M_2|} \sin2(\psi_{2}-\psi_{1}),
	\end{equation}
	where $\psi_{1}$ and $\psi_{2}$  are the phases associated with $a$ and $b$, respectively.  
	Similarly, using Eqns.(\ref{dirac}), (\ref{type2}) and Eqn.(\ref{epsilonD}) we have
	\begin{equation}\label{epsilonD1}
	\epsilon_{\Delta}=\frac{ M_1}{8\pi v_{H}^{2}}(3 |d| \sin(\psi_{d}-2\psi_{1})+3 |f| \sin(\psi_{f}-2\psi_{1})),
	\end{equation} 
	where $\psi_{d}$ and $\psi_{f}$ are the phases associated with parameter $d$ and $f$, respectively. Interestingly, while the type-II contribution to neutrino mass matrix in Eqn.(\ref{type2}) contains three model parameters $d$, $p$ and $f$ but the $CP$ asymmetry $\epsilon_{\Delta}$ contains $|d|$ and $|f|$ only, thus, reducing the number of model parameters responsible for producing net $CP$ asymmetry. This feature of the model is due to the structure of $m_{\nu_2}$ in Eqn.(\ref{type2}) wherein the sum of elements of row/column is independent of parameter $p$. The scalar triplet taking part in type-II seesaw contributes to the $CP$ asymmetry as well as, is responsible for $\mu-\tau$ symmetry breaking, thus, results in non-zero $\theta_{13}$.

	\begin{table}
		\centering
		\begin{tabular}{llc}
			
			Parameters  &  Best-fit$\pm 1\sigma$ & 3$\sigma$ range\\
			\hline
			$\Delta m_{21}^2[10^{-5}$ eV$^2$] & $7.55^{+0.20}_{-0.16}$ & $7.05-8.14$ \\
			$\Delta m_{31}^2[10^{-3}$ eV$^2$](NH) & $2.50\pm0.03$ & $2.41-2.60$ \\
			$\Delta m_{31}^2[10^{-3}$ eV$^2$](IH)& $2.42^{+0.03}_{-0.04}$ & $2.31-2.51$ \\
			Sin$^2 \theta_{12}/10^{-1}$ & $3.20^{+0.20}_{-0.16}$ & $2.73-3.79$ \\
			Sin$^2 \theta_{23}/10^{-1}$(NH) & $5.47^{+0.20}_{-0.30}$ & $4.45-5.99$ \\
			Sin$^2 \theta_{23}/10^{-1}$(IH) & $5.51^{+0.18}_{-0.30}$ & $4.53-5.98$ \\
			Sin$^2 \theta_{13}/10^{-2}$(NH) & $2.160^{+0.083}_{-0.069}$ & $1.96-2.41$ \\
			Sin$^2 \theta_{13}/10^{-2}$(IH) & $2.220^{+0.074}_{-0.076}$ & $1.99-2.44$ \\
		\end{tabular}
		\caption{\label{tab2}The latest global data on neutrino oscillation parameters used in the numerical analysis\cite{data}.}
	\end{table}

	\section{Numerical Analysis and Results }
	In order to find the $CP$ asymmetry parameter $\epsilon$, we need to find values of the model parameters. To attain this we compare the neutrino mass matrix obtained from the $A_4$ model(Eqn.(\ref{total})) with the low energy effective mass matrix.
	\par In $m_l$-diagonal basis, the effective neutrino mass matrix is given by 
	
	\begin{equation}\label{mpmns}
	M_{\nu} = V^{*}M_{\nu}^{diag}V^{\dagger},
	\end{equation}
	where $V=U.P$ and $M_{\nu}^{diag} = diag(m_1,m_2,m_3)$. $U$ is the Pontecorvo–Maki–Nakagawa–Sakata(PMNS) mixing matrix given by
	\begin{equation}
	U={
		\begin{pmatrix}
		c_{12}c_{13}  & s_{12} c_{13} & s_{13}e^{-i\delta}  \\
		-s_{12}c_{23}-c_{12}s_{13}s_{23} e^{i\delta} & c_{12}c_{23}-s_{12}s_{13}s_{23}e^{i\delta}& c_{13}s_{23}  \\
		s_{12} s_{23}-c_{12}s_{13}c_{23}e^{i\delta} & -c_{12}s_{23}-s_{12}s_{13}c_{23}e^{i\delta}& c_{13}c_{23}
		\end{pmatrix}},
	\end{equation}
	where $s_{kl}= \sin \theta_{kl}$ and $c_{kl}= \cos \theta_{kl}$ and $P$ is the diagonal phase matrix, $P= diag(1,e^{i\alpha},e^{i(\beta+\delta)})$, where $\delta$ is Dirac phase and $\alpha$, $\beta$ are two Majorana phases. In general, the effective neutrino mass matrix can be written as
	\begin{equation} \label{generalmnu}
	M_{\nu} = 
	{\begin{pmatrix}
		M_{11}& M_{12}&M_{13}\\
		M_{12}&M_{22}&M_{23} \\
		M_{13}& M_{23}&M_{33} 
		\end{pmatrix}},
	\end{equation}
	\begin{figure}[t]
	\begin{center}
		{\epsfig{file=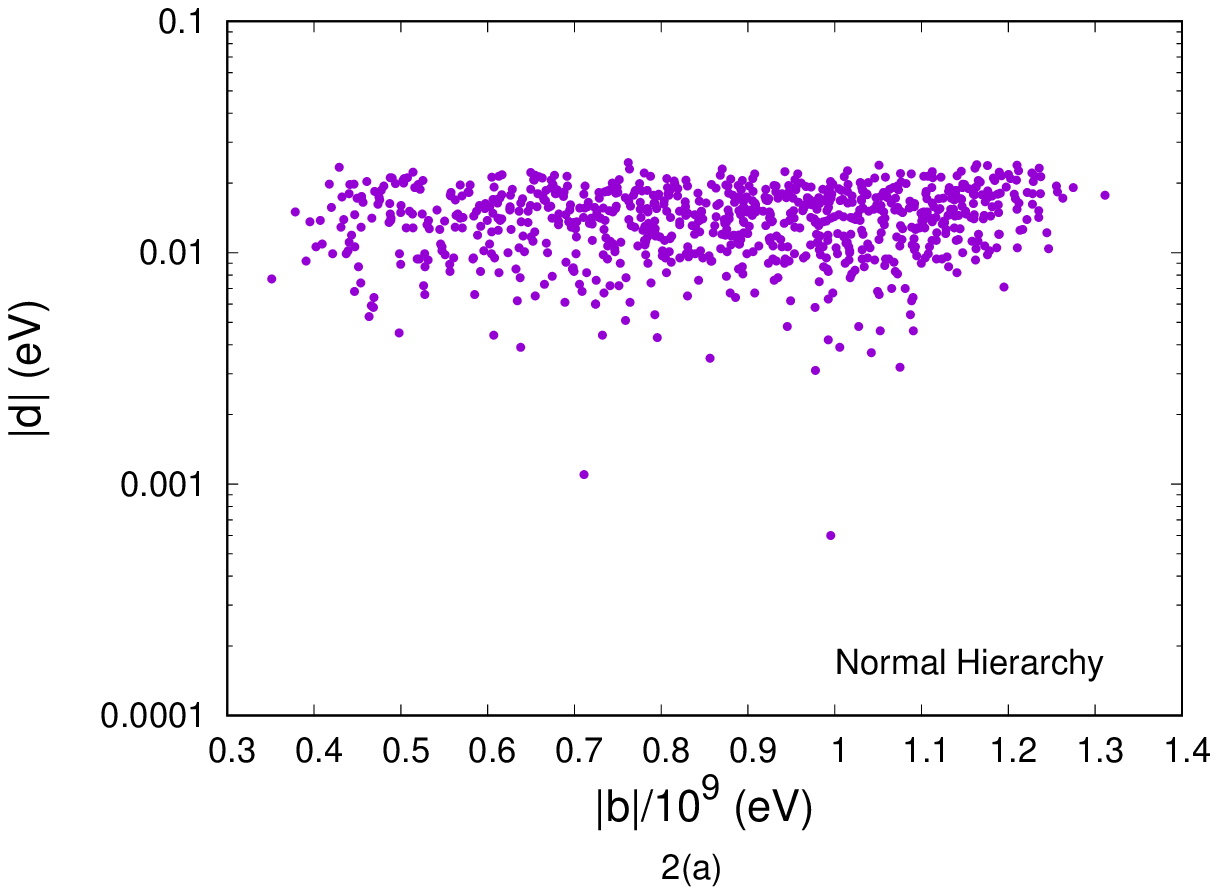,height=5.0cm,width=7.0cm}
			\epsfig{file=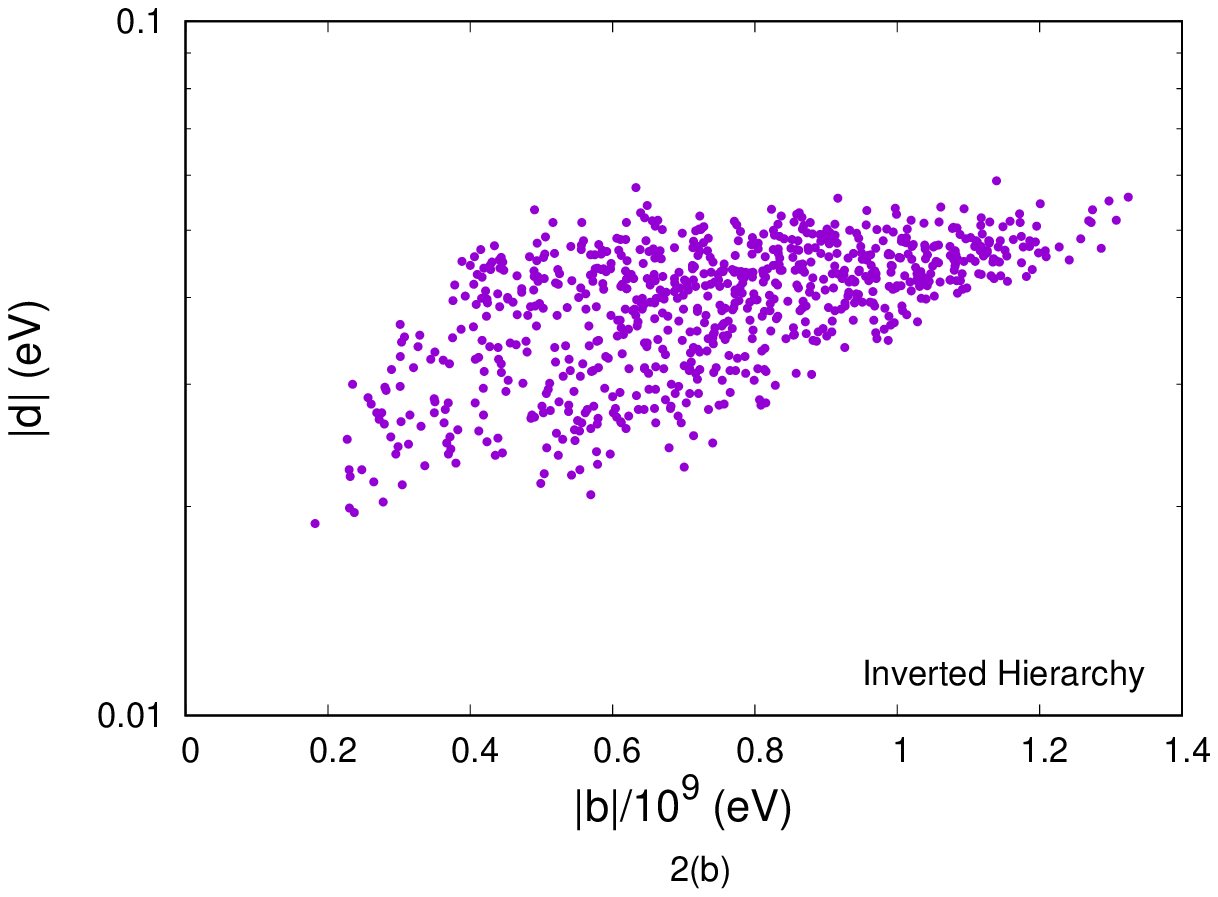,height=5.0cm,width=7.0cm}}\\
		{\epsfig{file=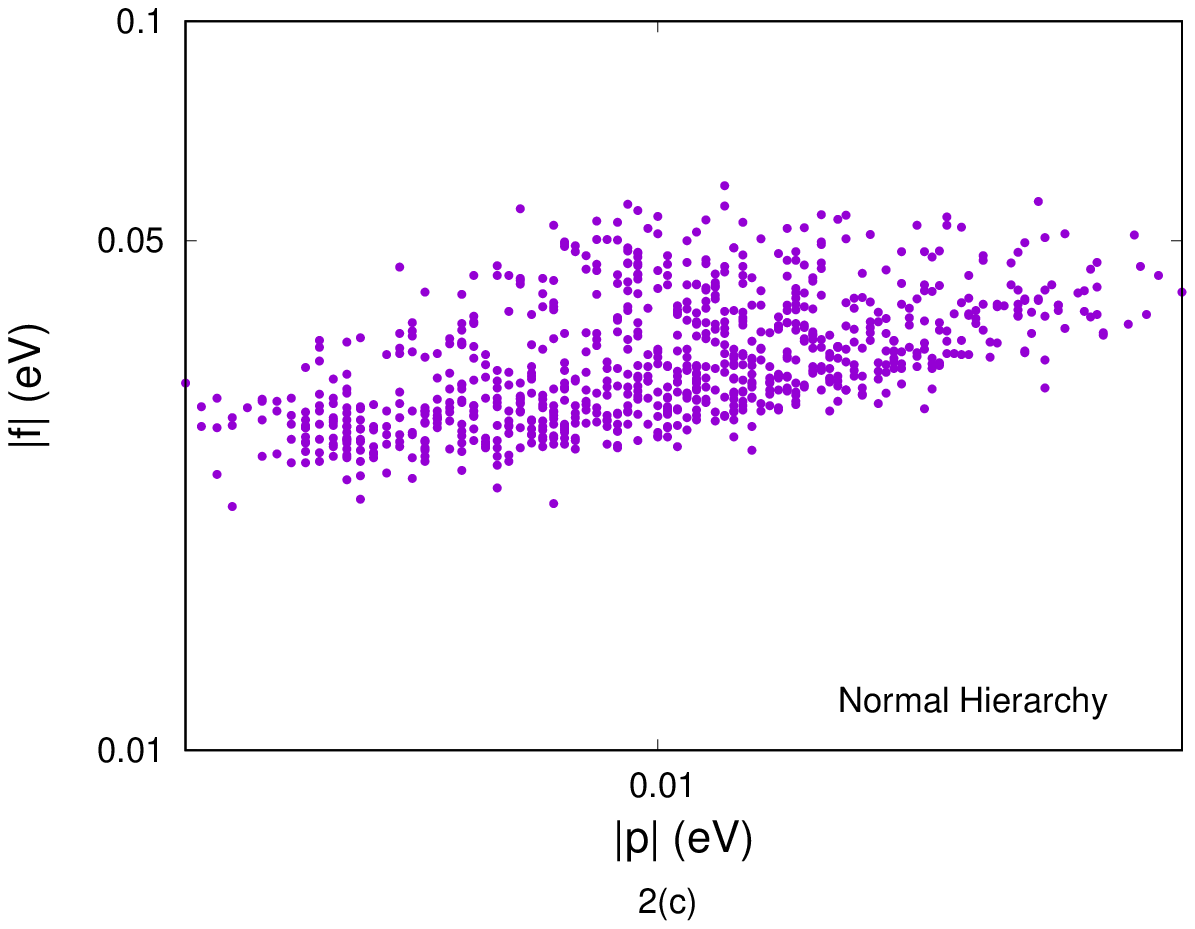,height=5.0cm,width=7.0cm}
			\epsfig{file=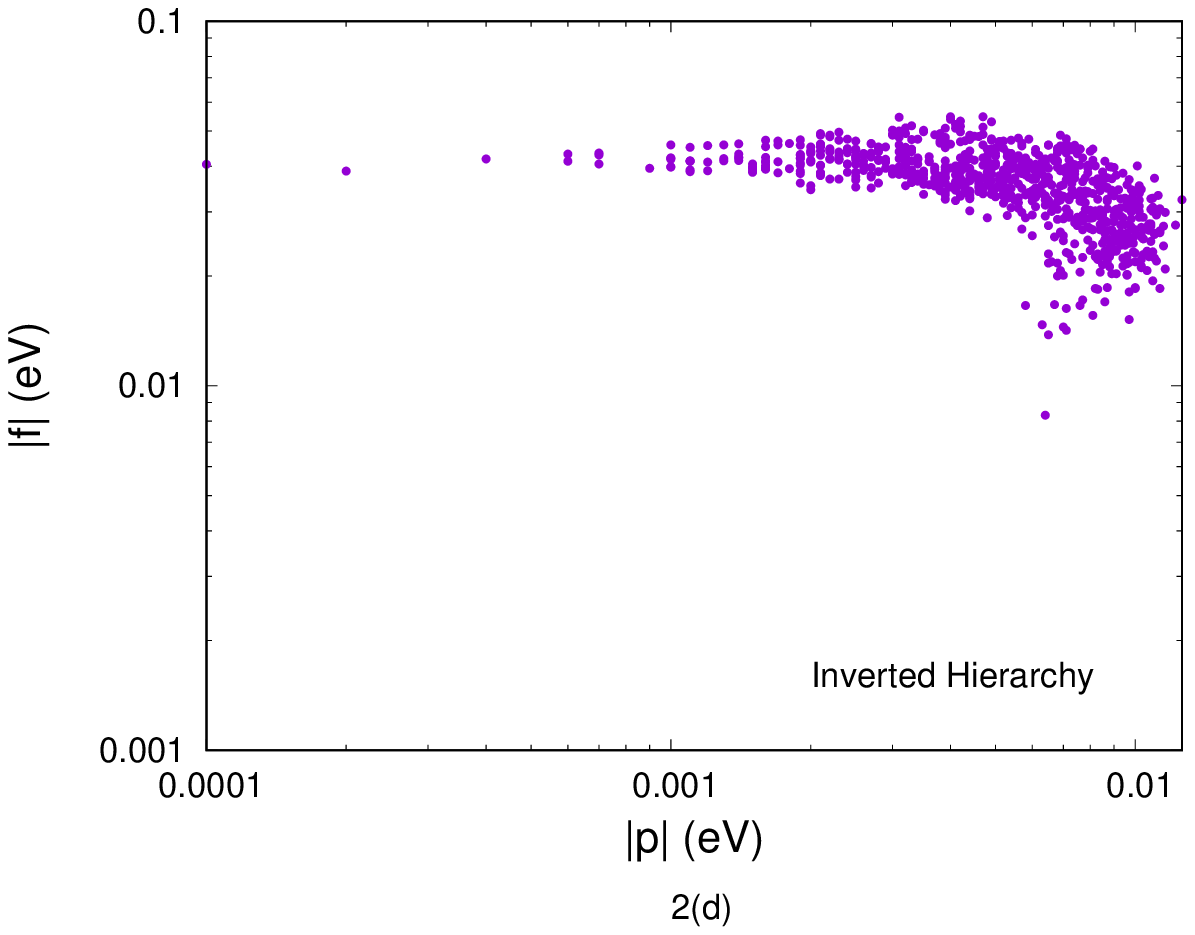,height=5.0cm,width=7.0cm}}
	\end{center}
	\caption{~\label{fig2} The correlation plots amongst the model parameters $|b|,|d|$ and $|f|$ at 3$\sigma$ C.L..}
\end{figure}

\begin{figure}[t]
	\begin{center}
		{\epsfig{file=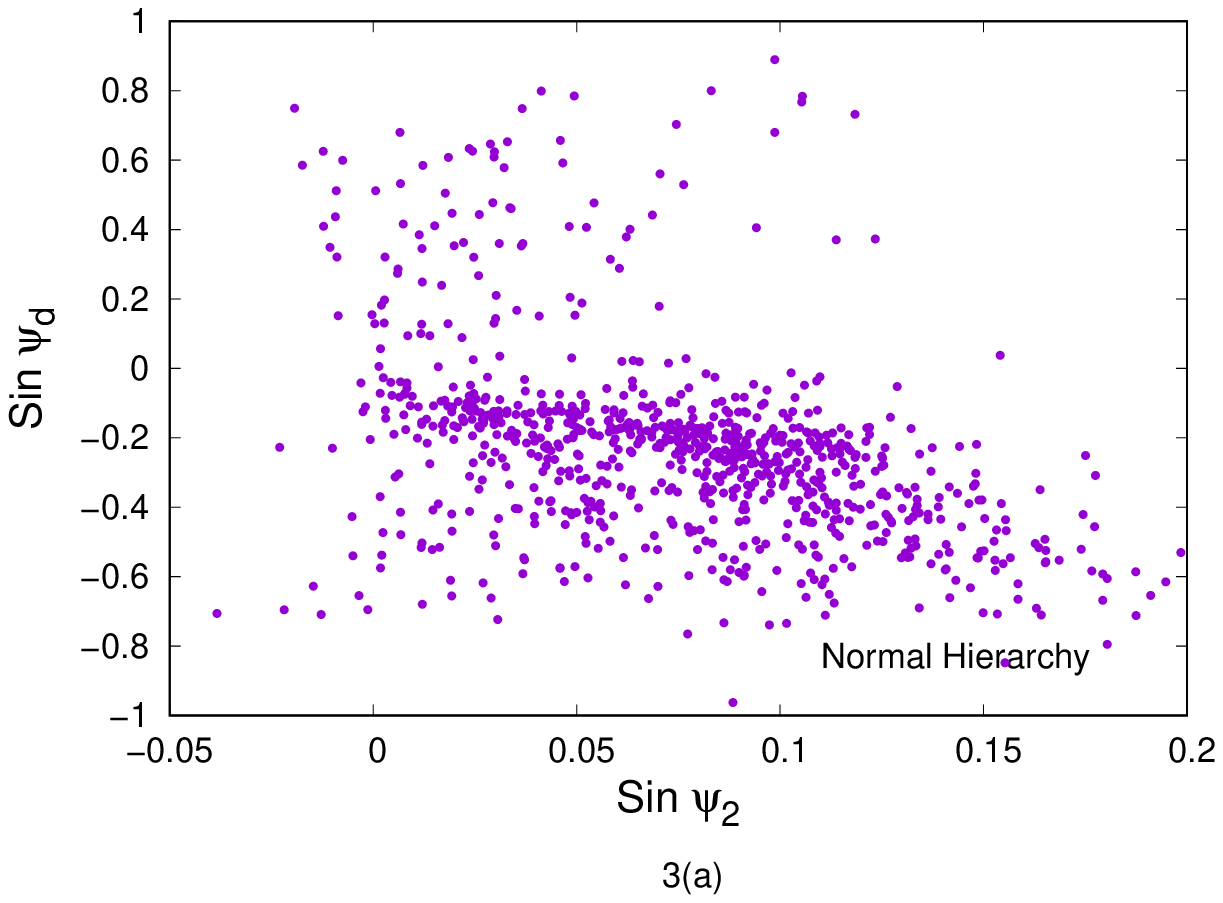,height=5.0cm,width=7.0cm}
			\epsfig{file=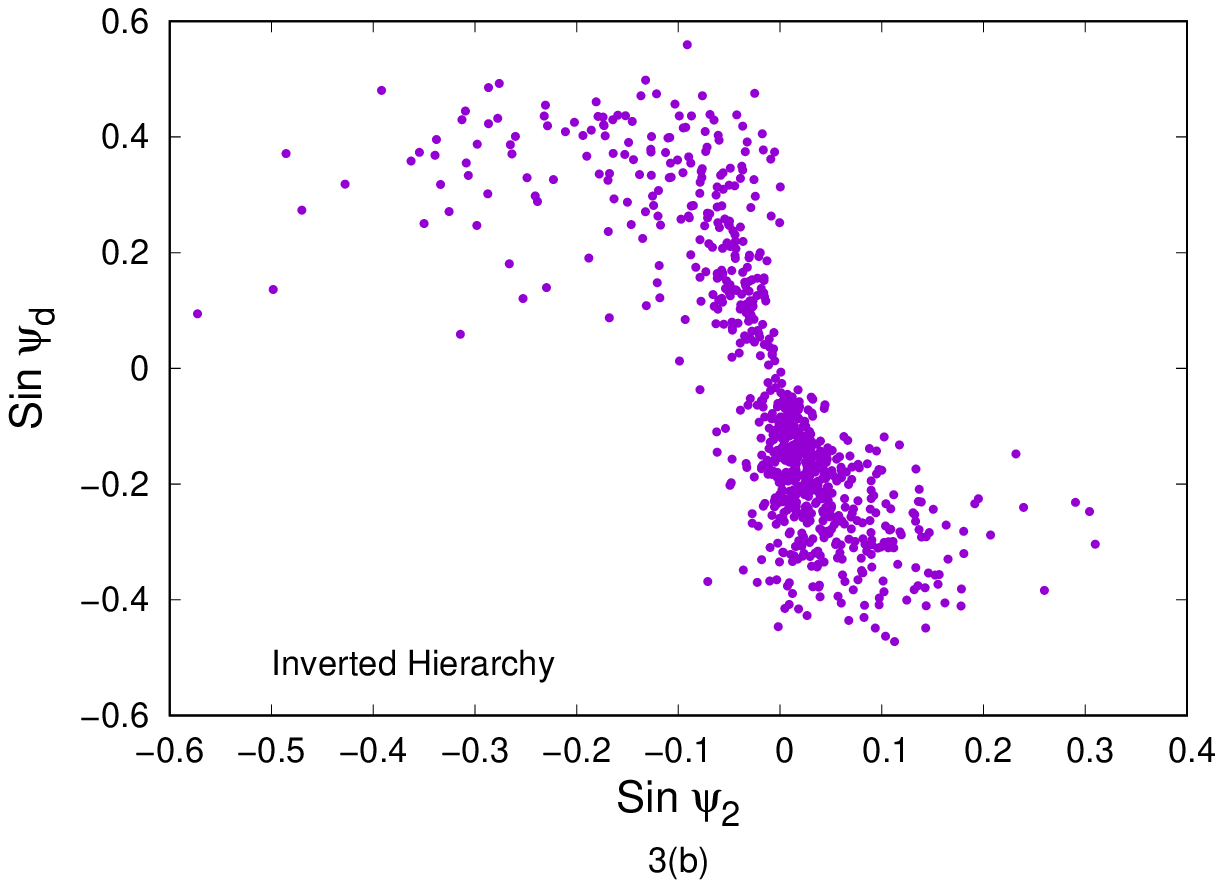,height=5.0cm,width=7.0cm}}\\
		{\epsfig{file=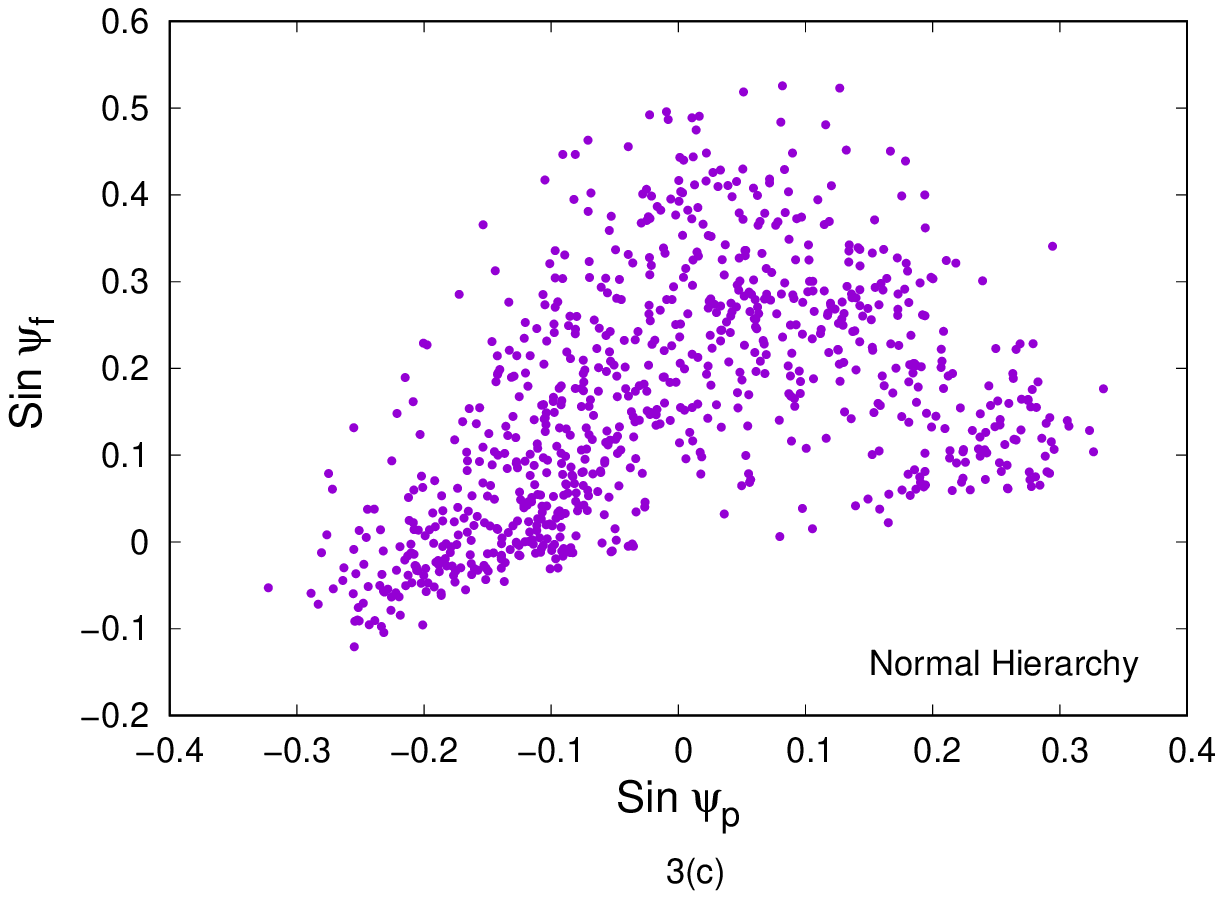,height=5.0cm,width=7.0cm}
			\epsfig{file=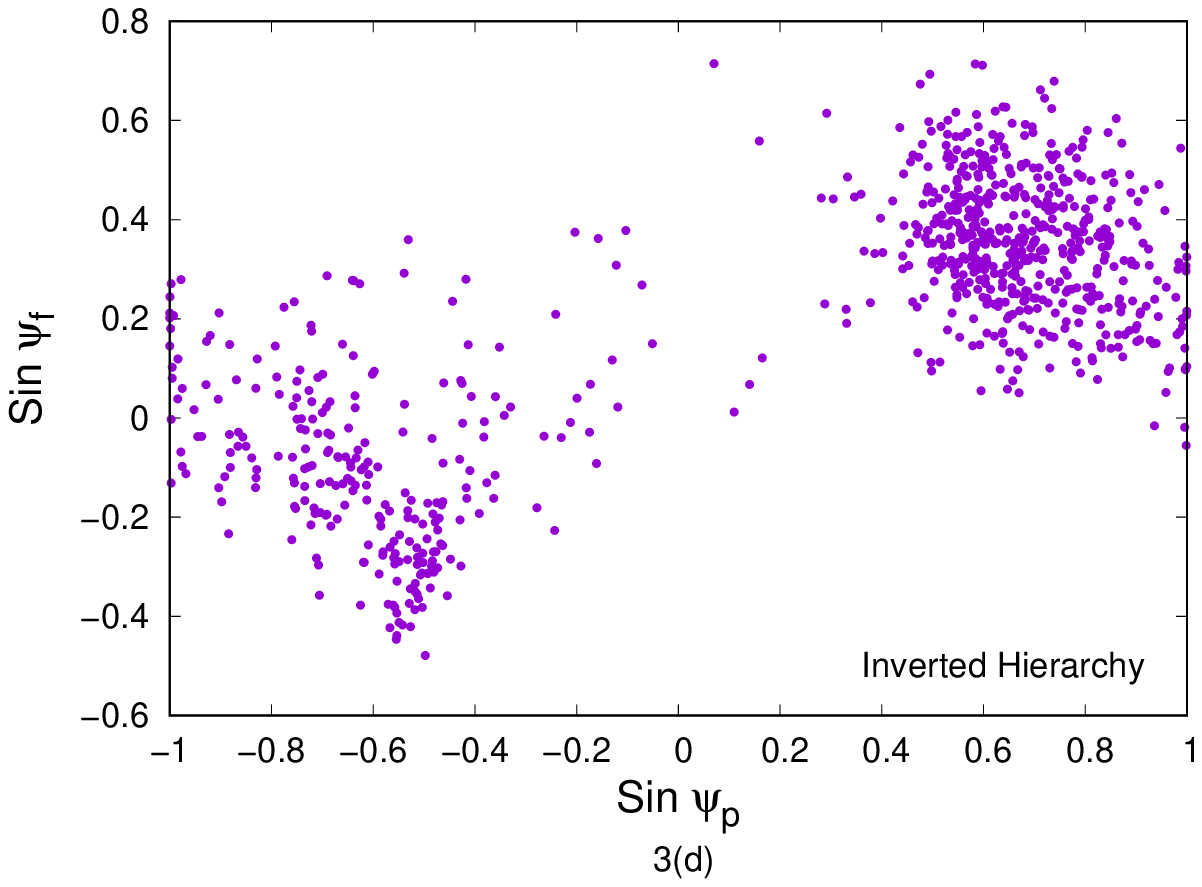,height=5.0cm,width=7.0cm}}
	\end{center}
	\caption{~\label{fig3} The correlation plots amongst the phases $\psi_{2},\psi_{d}, \psi_{p}$ and $\psi_{f}$ at 3$\sigma$ C.L..}
\end{figure}

\begin{figure}[t]
	\begin{center}
		{\epsfig{file=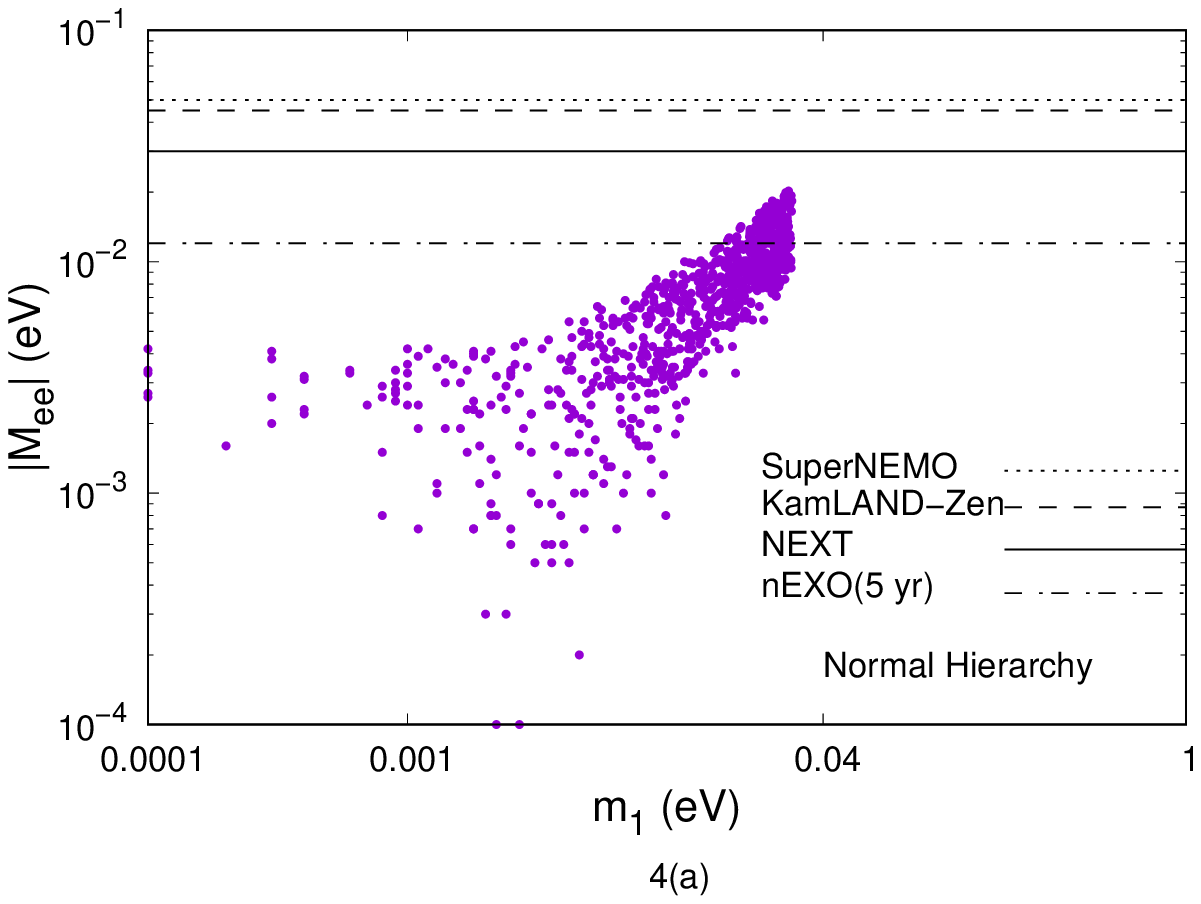,height=5.0cm,width=7.0cm}
			\epsfig{file=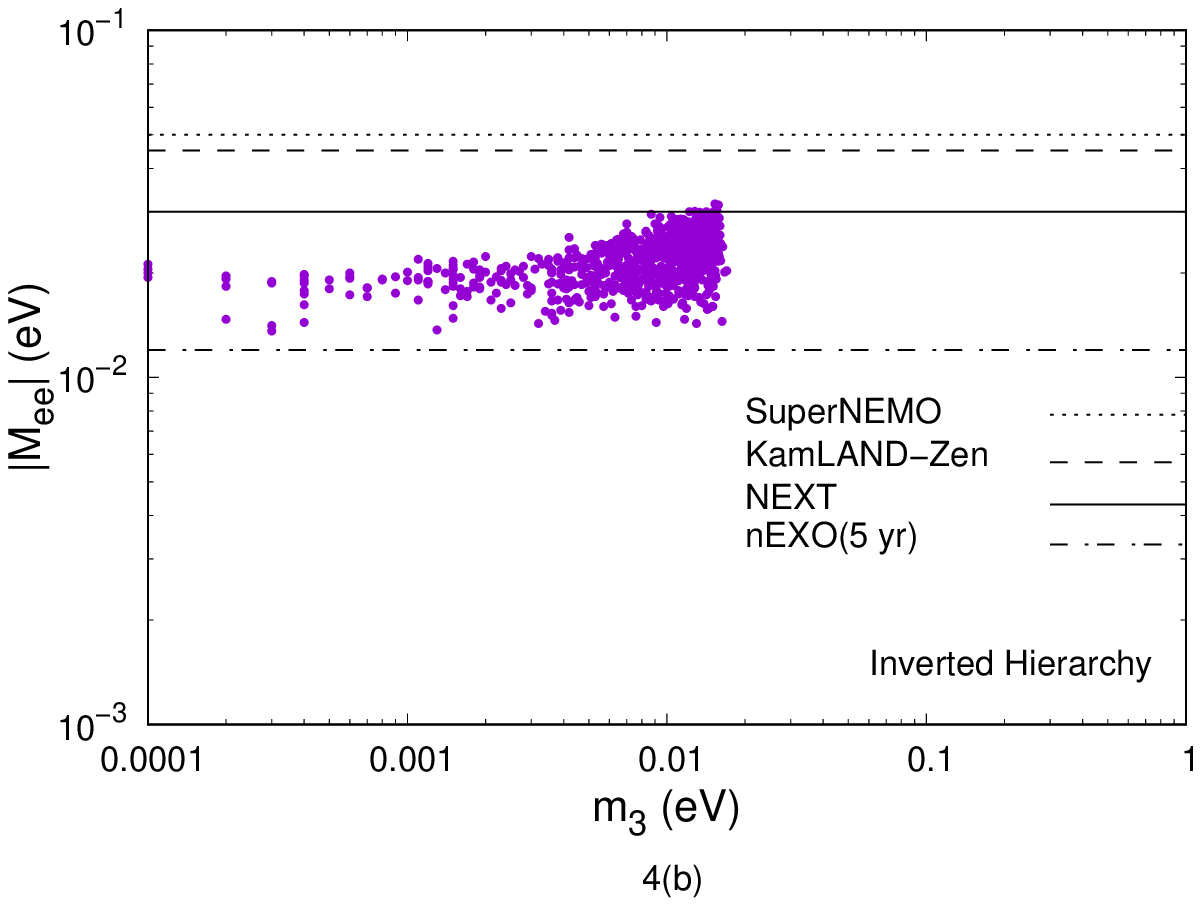,height=5.0cm,width=7.0cm}}\\
		{\epsfig{file=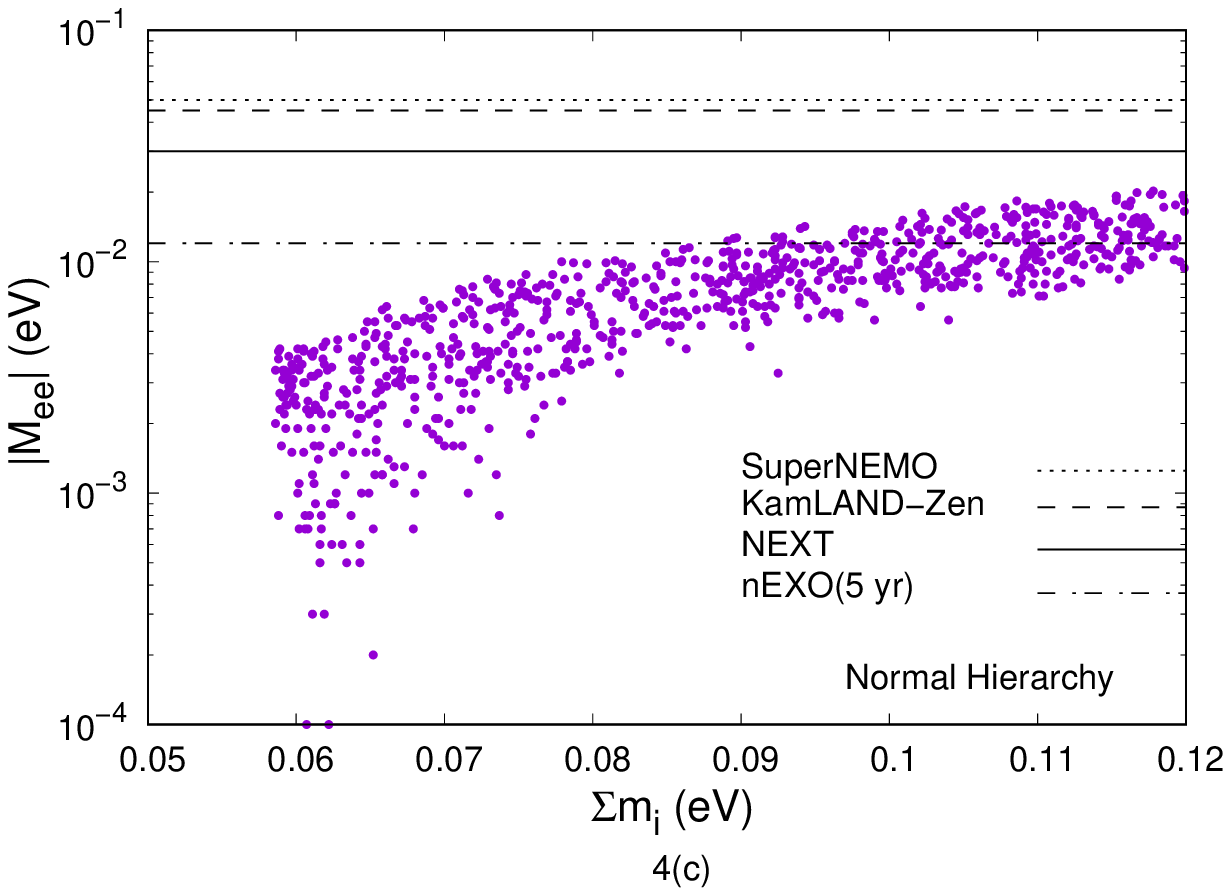,height=5.0cm,width=7.0cm}
			\epsfig{file=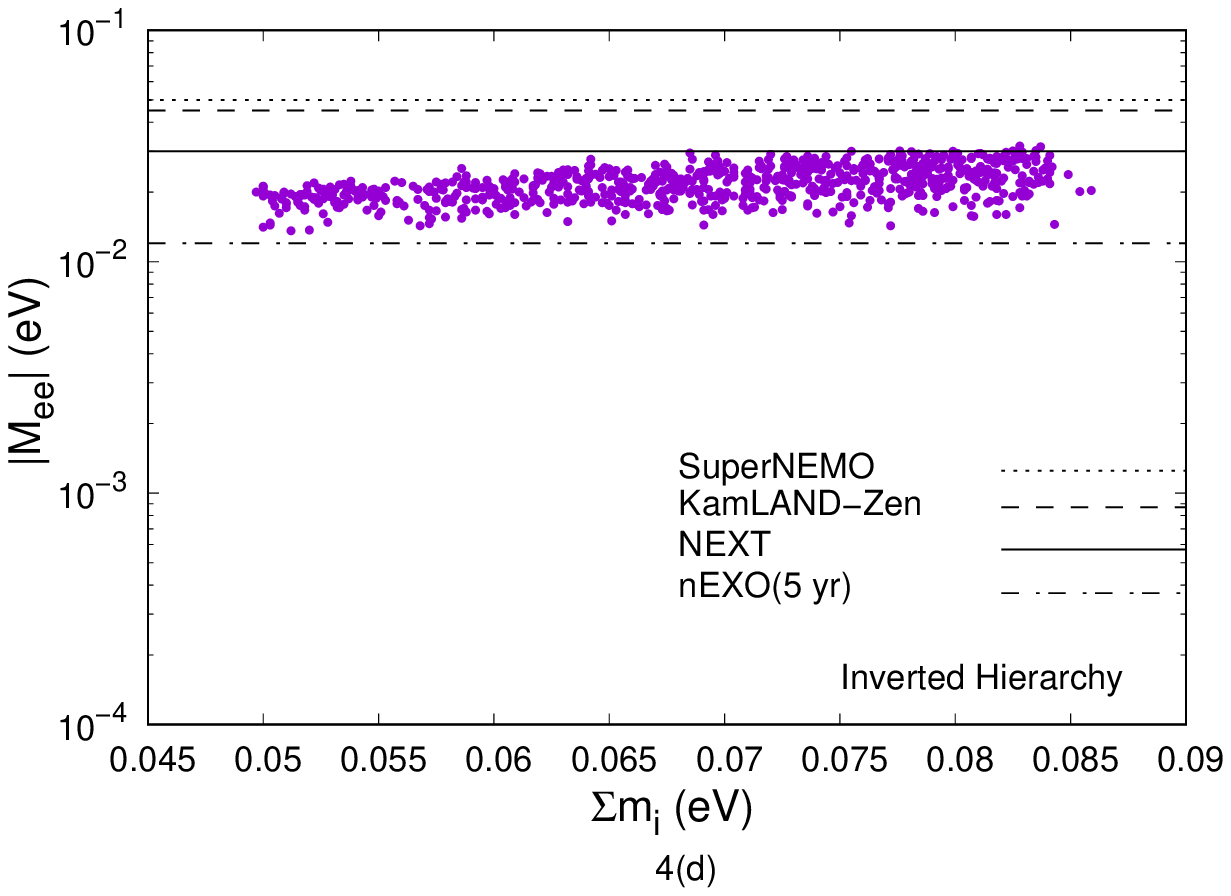,height=5.0cm,width=7.0cm}}
	\end{center}
	\caption{~\label{fig4} The variation of effective Majorana neutrino mass $|M_{ee}|$ with lightest neutrino mass(first row) and with sum of absolute neutrino masses $\Sigma m_{i}$(second row) for normal as well as inverted hierarchy at 3$\sigma$ C.L.. The current sensitivity limits of various 0$\nu\beta\beta$-decay experiments are, also, shown.}
\end{figure}
where $M_{ij}$ (for $ i,j=1,2,3$) contains nine parameters viz., three neutrino masses $m_1,m_2,m_3$, three mixing angles $\theta_{12},\theta_{23},\theta_{13}$, and three $CP$ violating phases $\delta,\alpha,\beta$. More explicitly,
\begin{equation}\label{elements}
    \begin{rcases}
M_{11}&=c_{13}^2 (m_1c_{12}^2+ m_2 s_{12}^2e^{-2i\alpha}) + m_3 e^{-2i\beta}s_{13}^2,\\ 
M_{12}&= c_{13} (e^{-i(2 \beta + \delta)} m_3 s_{13} s_{23} - 
   c_{12} m_1 (c_{23} s_{12} + c_{12} e^{-i\delta} s_{13} s_{23})\\
   &+ 
   e^{-2i\alpha} m_2 s_{12} (c_{12} c_{23} - e^{-i\delta} s_{12} s_{13} s_{23})),\\ 
   M_{13}&=c_{13} (c_{23} e^{-i(2\beta + \delta)} m_3 s_{13} - 
e^{-2i\alpha} m_2 s_{12} (c_{23} e^{-i\delta} s_{12} s_{13}+ c_{12} s_{23})\\ 
&+ 
c_{12} m_1 (-c_{12} c_{23} e^{-i\delta} s_{13} + s_{12} s_{23})),   \\ 
M_{22}&=c_{13}^2 e^{-2i(\beta + \delta)} m_3 s_{23}^2 + 
 m_1 (c_{23} s_{12} + c_{12} e^{-i\delta} s_{13} s_{23})^2+ 
 e^{-2i\alpha} m_2 (c_{12} c_{23} - e^{-i\delta} s_{12} s_{13} s_{23})^2,\\
 M_{23}&=c_{13}^2 c_{23} e^{-2i(\beta + \delta)} m_3 s_{23} + 
 m_1 (c_{12} c_{23} e^{-i\delta} s_{13} - s_{12} s_{23}) (c_{23} s_{12} + 
    c_{12} e^{-i\delta}s_{13} s_{23})\\
    &-e^{2i\alpha}m_2 (c_{23} e^{-i\delta} s_{12} s_{13} + c_{12} s_{23}) (c_{12} c_{23} - 
    e^{-i\delta} s_{12} s_{13} s_{23}),\\
    M_{33}&= c_{13}^2 c_{23}^2 e^{-2i(\beta + \delta)} m_3 + 
 e^{-2i\alpha} m_2 (c_{23} e^{-i\delta} s_{12} s_{13} + c_{12} s_{23})^2 + 
 m_1 (c_{12} c_{23} e^{-i\delta} s_{13} - s_{12} s_{23})^2.
 \end{rcases}
 \end{equation} 

A comparison of Eqns.(\ref{total}) and (\ref{generalmnu}) yield the following relations amongst the model parameters ($c,p,d,f$) and the elements of the low energy neutrino mass matrix $M_{\nu}$

\begin{equation}\label{rela1}
\begin{rcases}
c&=\frac{2 M_{23}+M_{11}}{3}, \\
p&=\frac{M_{11}-M_{23}}{3},\\
d&= M_{12}-M_{23},\\
f&=M_{22}-M_{11},
\end{rcases}
\end{equation}
also
\begin{equation}\label{rela2}
\begin{rcases}
M_{13}&=c-p+f,\\
M_{33}&=c+2p+d.
\end{rcases}
\end{equation}

Using the low energy data on neutrino oscillation parameters and full physical range(s) of $CP$ phases, Eqn.(\ref{rela1}) can be used to obtain the allowed parameter space of $c,p,d$ and $f$. For the framework to be self-consistent the elements M$_{13}$ and M$_{33}$ obtained from Eqn.(\ref{rela2}) as well as Eqn.(\ref{elements}) must be same within defined tolerance and acts as a necessary and sufficient condition to satisfy magic symmetry.

In the numerical analysis, the known neutrino oscillation parameters such as mixing angles ($\theta_{12}$, $\theta_{23}$, $\theta_{13}$) and mass-squared differences ($\Delta m_{21}^{2}$, $\Delta m_{31}^{2}$) are randomly generated using Gaussian distribution with the best-fit and errors given in Table \ref{tab2}. The unknown parameters viz., three $CP$ phases ($\delta,\alpha,\beta$) are randomly generated in their full physical range, ($0^o-360^o$), with uniform distribution.  The remaining unknown parameter i.e. absolute mass-scale $m_{lightest}$ ($m_{lightest}=m_1$ in normal hierarchy and $m_{lightest}=m_3$ in inverted hierarchy), is generated within a conservative range ($0-0.04$)eV guided by the cosmological limit on the sum of neutrino masses $\sum_{i} m_{i} < 0.12$eV (at 95 \% C.L.)\cite{cosmo}. The sample size for all these parameters constitutes of $10^7$ points. The neutrino masses $m_2$ and $m_3$($m_1$ and $m_2$) are calculated using the relations 

\begin{eqnarray}
\nonumber
m_2=\sqrt{m_1^{2}+\Delta m_{21}^{2}},m_3=\sqrt{m_1^{2}+\Delta m_{31}^{2}} \hspace{0.3cm}\text{for NH},\\
\nonumber
\end{eqnarray}
and
\begin{eqnarray}
\nonumber
m_1=\sqrt{m_{3}^{2}-\Delta m_{31}^{2}} ,m_2=\sqrt{m_3^{2}-\Delta m_{31}^{2}+\Delta m_{21}^{2}} \hspace{0.3cm}\text{for IH}.\\
\nonumber
\end{eqnarray}
The sum of $i^{th}$ row of neutrino mass matrix $M_\nu$ can be written as
\begin{equation}\label{magic}
S_{i}=M_{i1}+M_{i2}+M_{i3},
\end{equation}
where $ i=1,2,3$ is row index. The magic symmetry of $M_{\nu}$ implies $S_{1}=S_{2}=S_{3}=$constant$(\kappa)$. Using the procedure outlined above, we require that $M_{\nu}$ is magic symmetric within a tolerance of $10^{-3}$ i.e. if
\begin{eqnarray}
\nonumber
|S_{1}-S_{2}| \le10^{-3} \hspace{0.3cm}\text{and}\hspace{0.3cm}|S_{2}-S_{3}| \le10^{-3}. 
\end{eqnarray}

The parameter space, thus obtained, is used to calculate the $CP$ asymmetry parameter $\epsilon$ for normal(NH) as well as inverted hierarchy(IH). The model predictions are obtained taking into account the Davidson-Ibarra bound on right-handed neutrino mass scale\cite{dibound}. Hence, we have assumed the lightest right-handed neutrino mass of the order of $10^{9}$GeV. For successful leptogenesis, the right-handed neutrinos must be non-degenerate which is ensured by higher mass scale($M_2$) for second right-handed neutrino. We have assumed $M_2$ to be of the order $10^{10}$GeV. Also, in order to fulfill the condition for out-of-equilibrium decay in Eqn.(\ref{planck}), we find that $\frac{(m_{D}^{\dagger}m_{D})_{11}}{M_1}=\frac{3 a^{2}}{M_{1}} < 1.08 \times10^{-3}$eV. Thus, it is judicious to consider the approximations, $\frac{a^{2}}{M_{1}}\approx 10^{-4}$ and $\psi_1=0$. Using the calculated parameter space of $c=\frac{a^{2}}{M_{1}}+\frac{b^{2}}{M_{2}}$  alongwith $\frac{a^{2}}{M_{1}}\approx10^{-4}$ and $\psi_1=0$, in Eqns.(\ref{eta}-\ref{epsilonD1}), we obtain the  parameter space of $b$ which contributes to the $CP$ asymmetry (Eqn.(\ref{epsilonN1})).

\begin{figure}
	\begin{center}
		{\epsfig{file=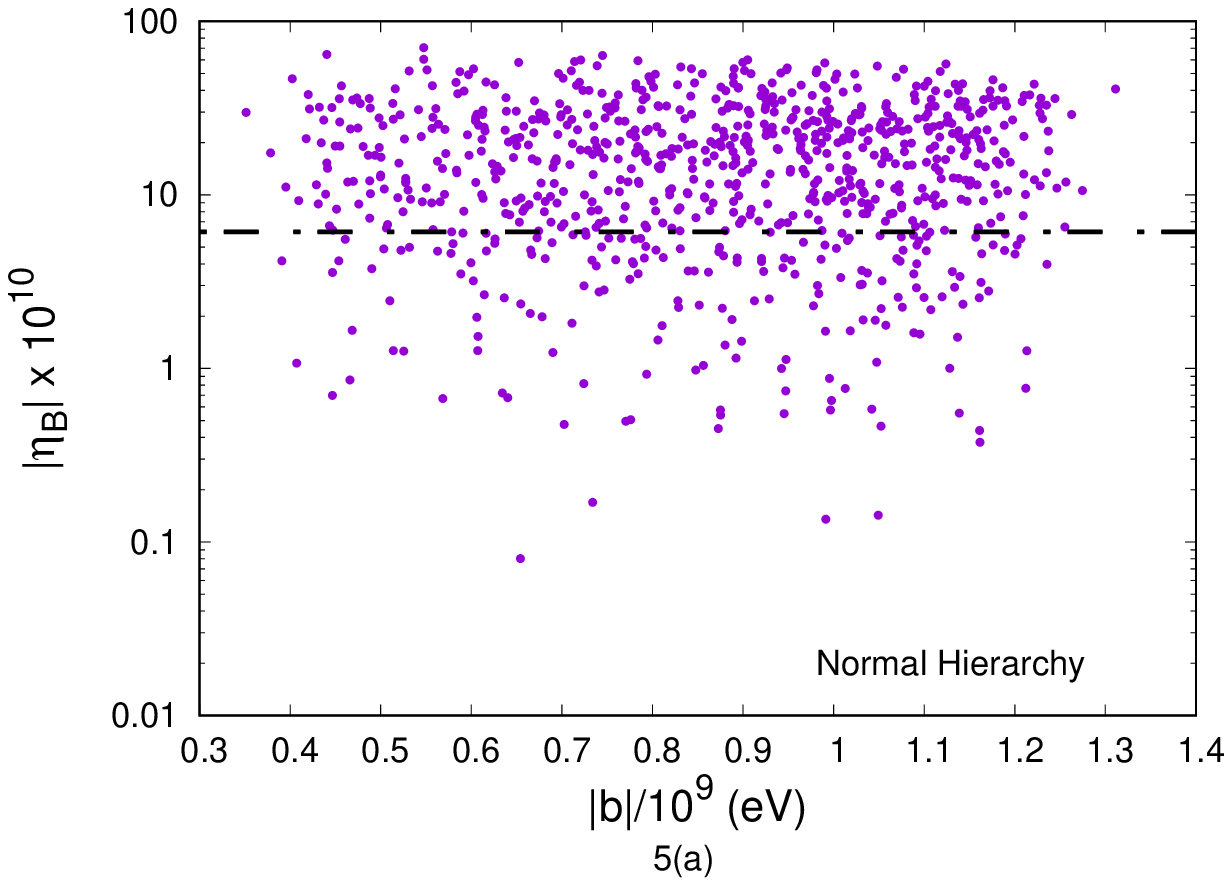,height=5.0cm,width=7.0cm}
			\epsfig{file=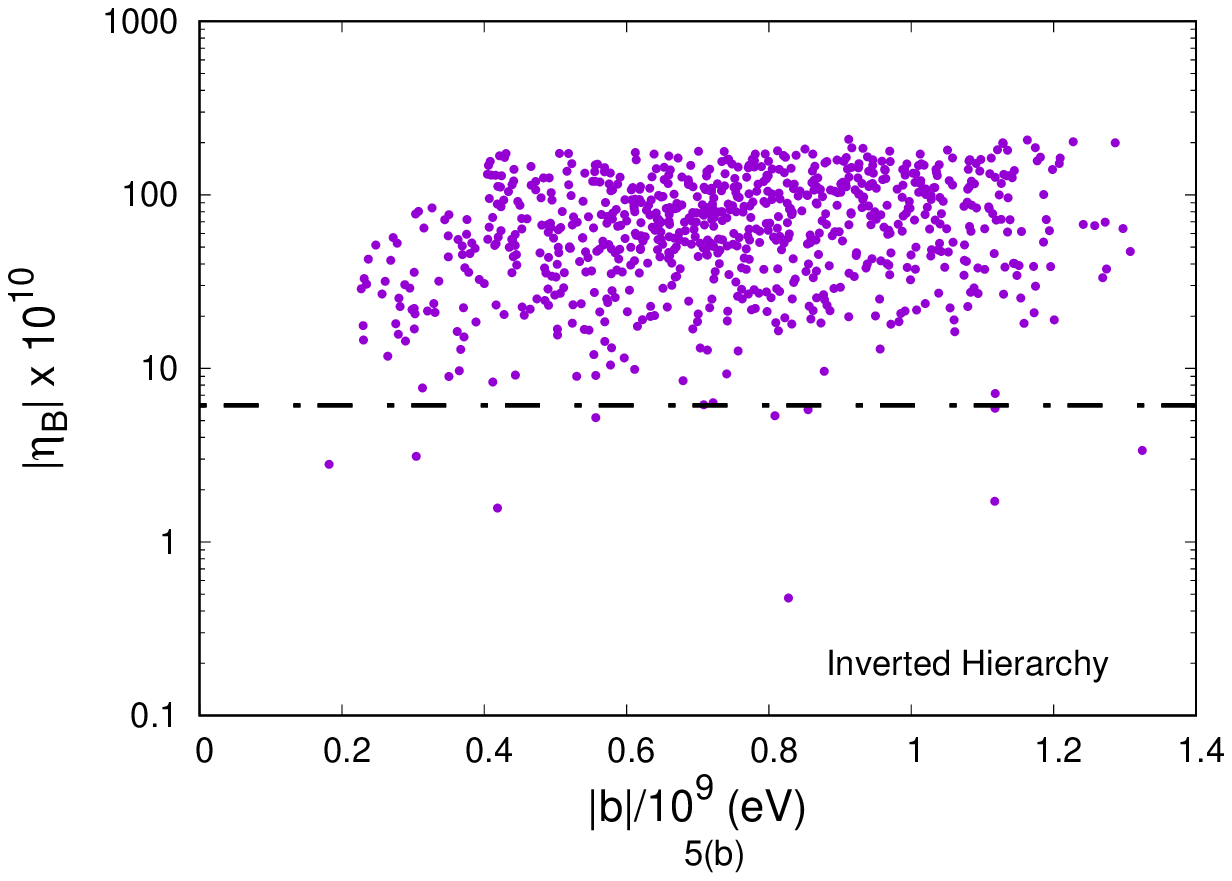,height=5.0cm,width=7.0cm}}\\
		{\epsfig{file=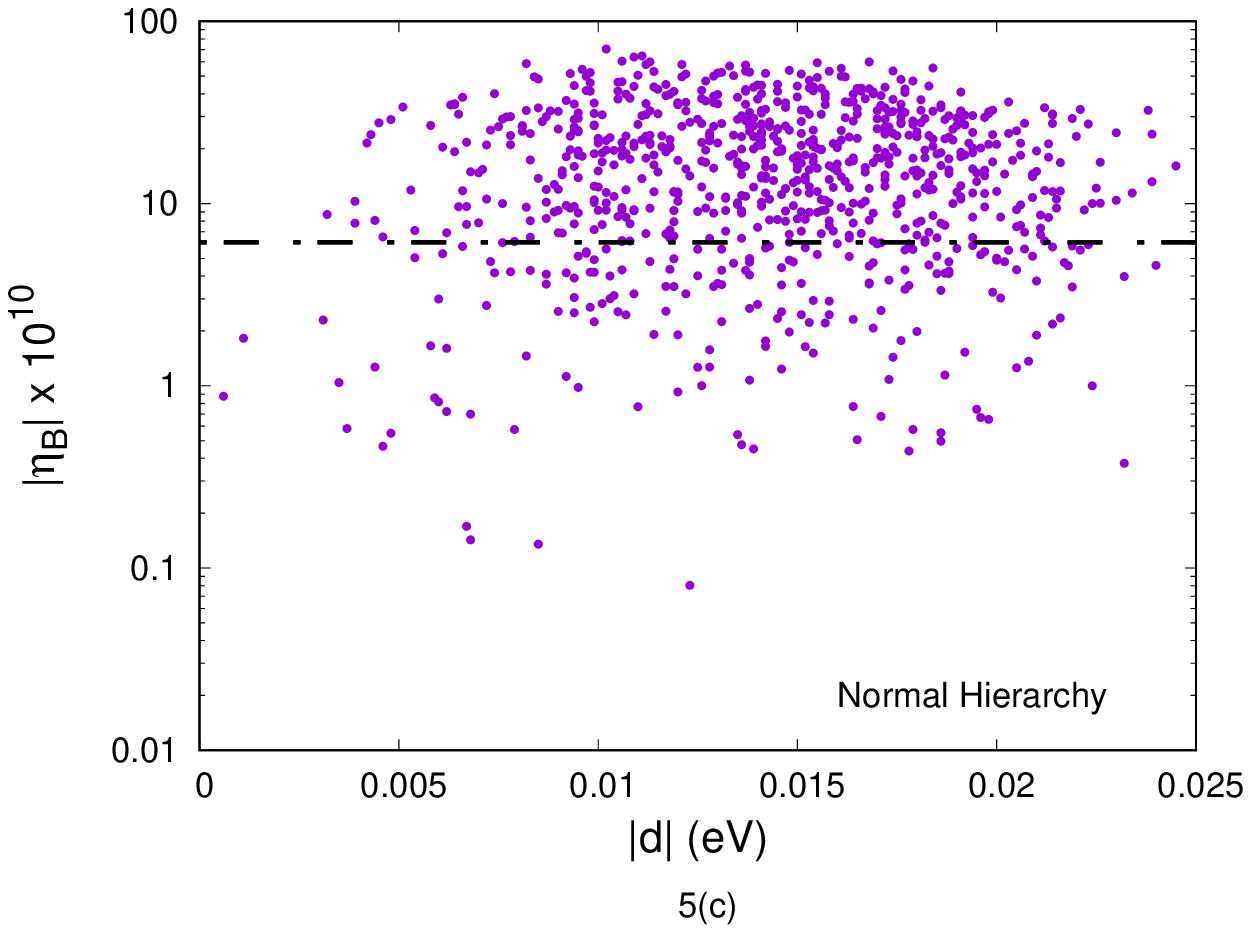,height=5.0cm,width=7.0cm}
			\epsfig{file=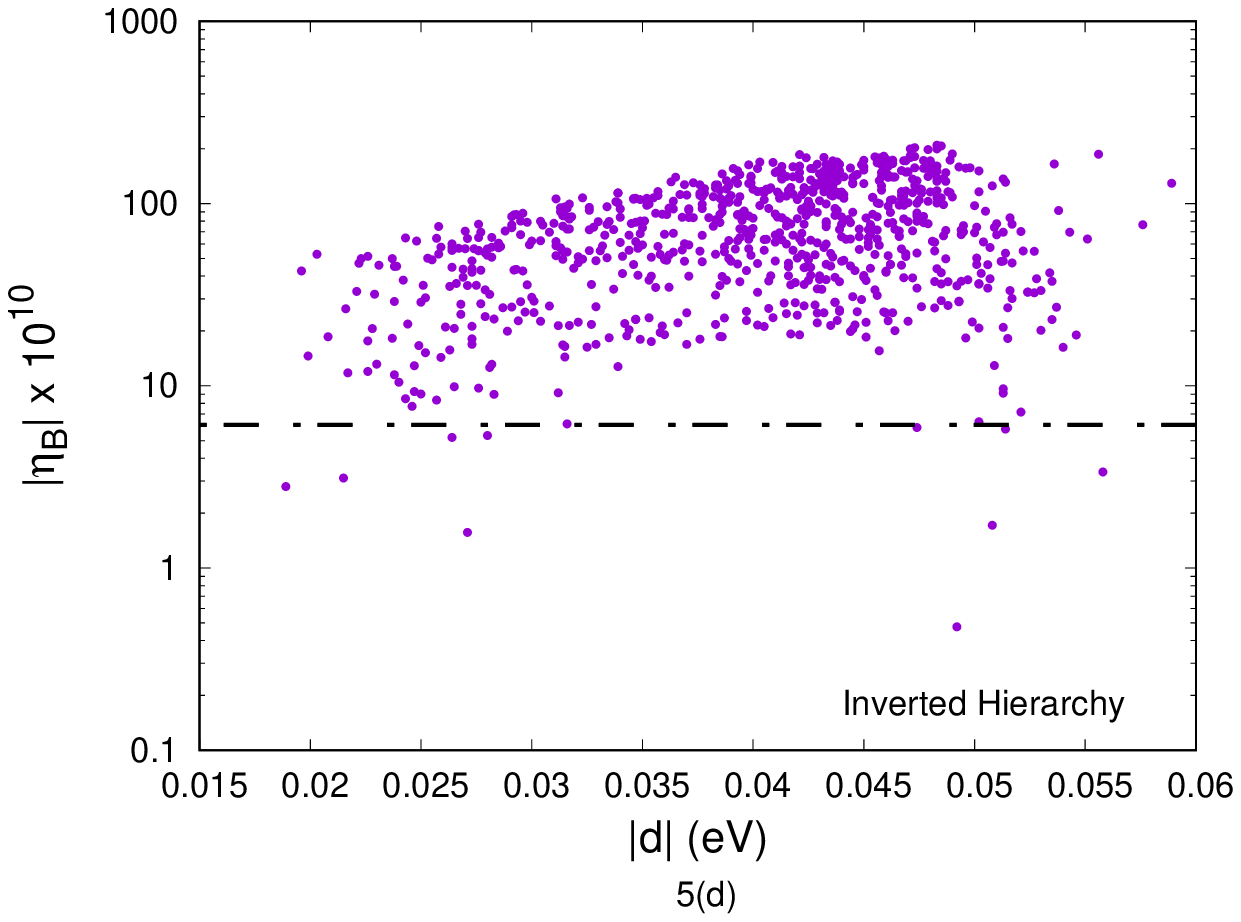,height=5.0cm,width=7.0cm}}\\
		{\epsfig{file=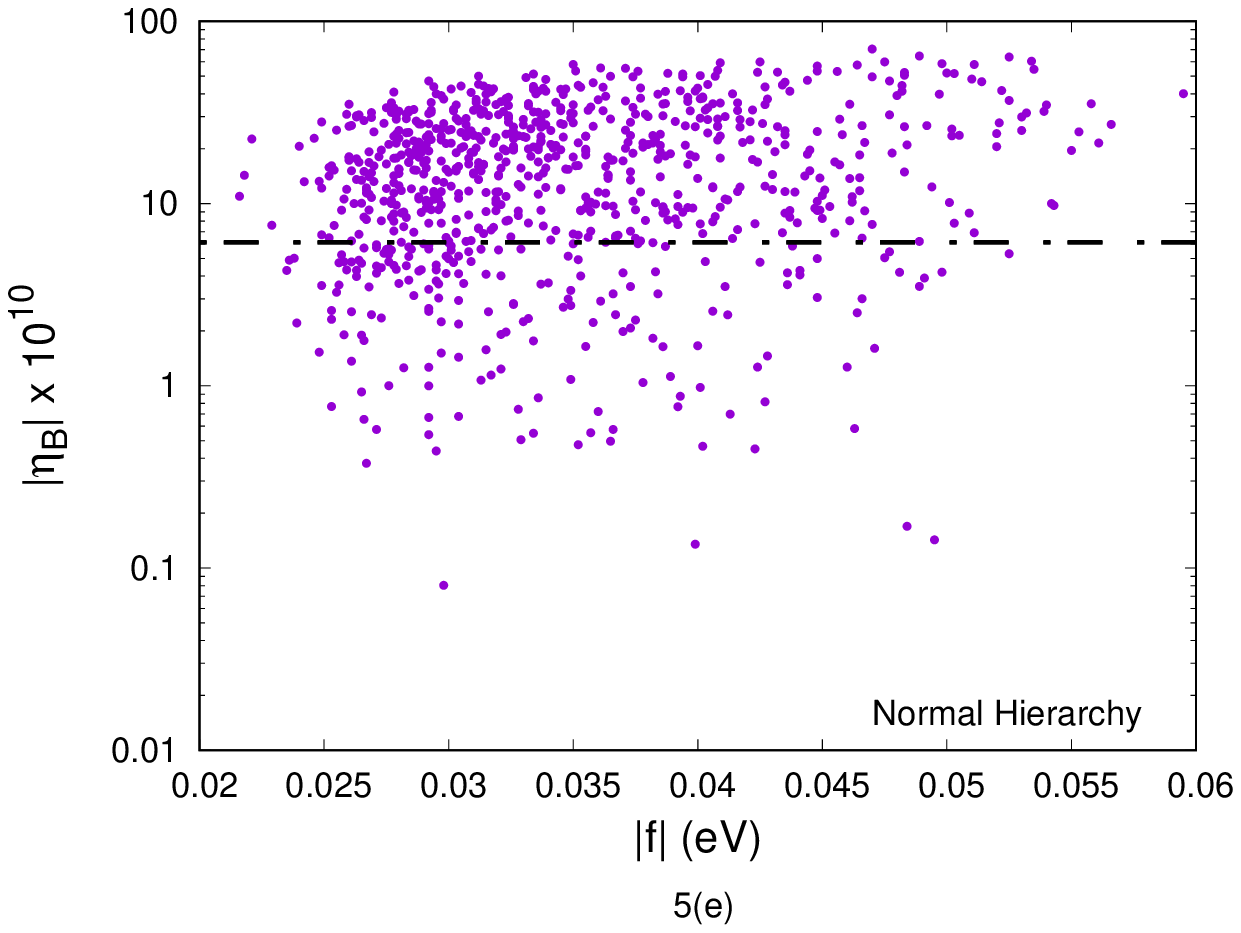,height=5.0cm,width=7.0cm}
			\epsfig{file=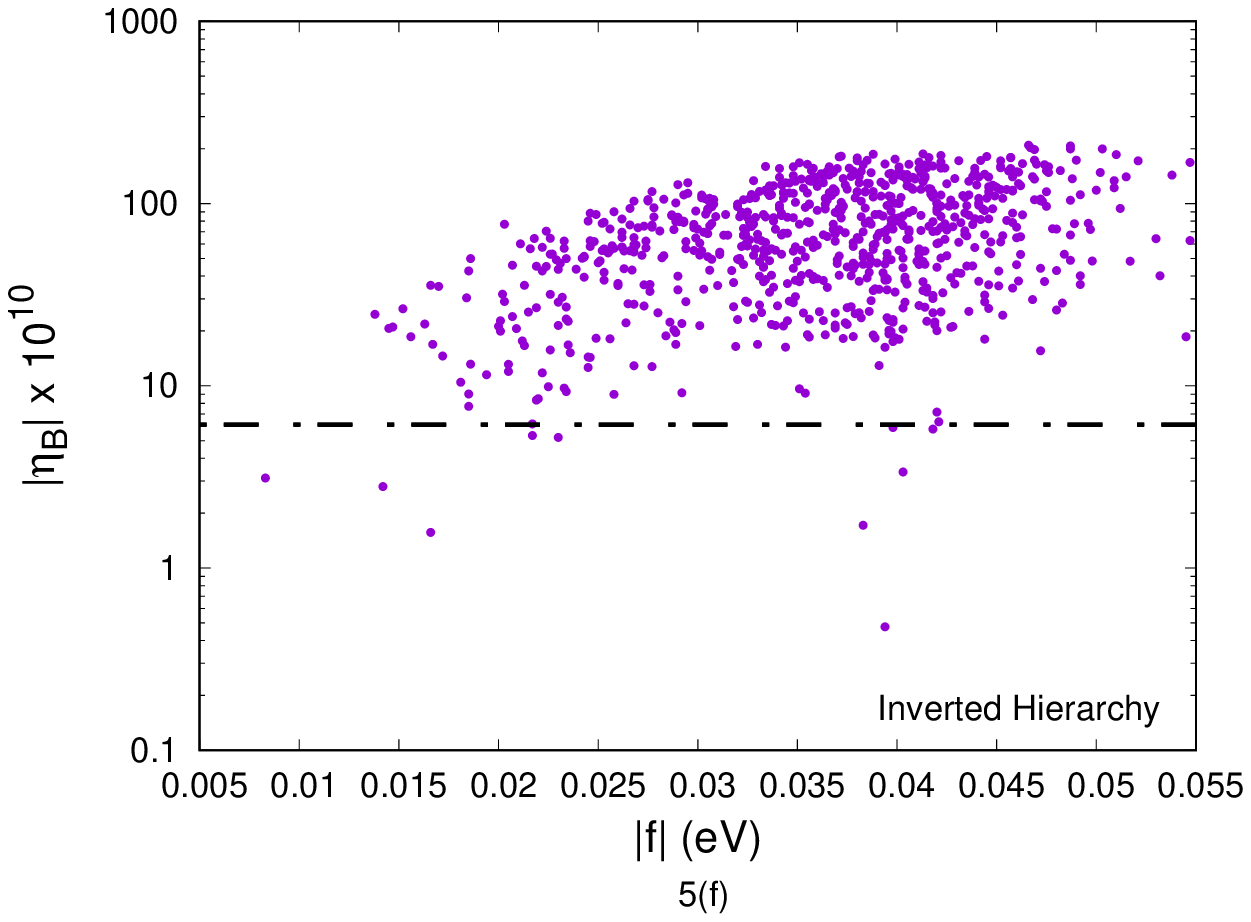,height=5.0cm,width=7.0cm}}\\
	\end{center}
	\caption{\label{fig5} The variation of baryon asymmetry with $|b|$(first row), $|d|$(second row) and $|f|$(third row) for normal and inverted hierarchy. The dot-dashed line is the observed baryon asymmetry $|\eta_B|=(6.12 \pm 0.04) \times 10^{-10}$.}
\end{figure}
The phases $\psi_{2}$, $\psi_p$, $\psi_{d}$ and $\psi_{f}$ associated with the model parameters $b$, $p$, $d$ and $f$, respectively, can be evaluated using the relations
\begin{equation}\label{phases}
\begin{rcases}
\psi_{2}&=\frac{1}{2}Arg\left({\frac{2 M_{23}+M_{11}}{3}}\right), \\
\psi_{p}&=Arg\left(\frac{M_{11}-M_{23}}{3}\right),\\
\psi_{d}&= Arg\left(M_{12}-M_{23}\right),\\
\psi_{f}&=Arg\left({M_{22}-M_{11}}\right).
\end{rcases}
\end{equation}

 	\par The type-II seesaw plays pivotal role to have correct neutrino phenomenology (non-zero $\theta_{13}$ and non-degenerate neutrino masses) and leptogenesis. In Fig.(\ref{fig2}) and Fig.(\ref{fig3}) we have shown the correlations amongst the model parameters($|b|$, $|p|$, $|d|$, $|f|$) and phases($\psi_{2},\psi_p,\psi_{d},\psi_{f}$), respectively, induced by the magic symmetric neutrino mass matrix and allowed by the current data on neutrino mass and mixings.
 	
	In the effective neutrino mass matrix (Eqn.(\ref{total})), $|(m_{\nu})_{11}|\equiv|M_{ee}|$ element of neutrino mass matrix is non-zero which results in non-vanishing effective Majorana mass. The variation of $|M_{ee}|$ with lightest neutrino mass and sum of light neutrino masses $\Sigma m_{i}$ is shown in Fig.(\ref{fig4}) for both the neutrino mass hierarchies(NH and IH). The sensitivity reach of 0$\nu\beta\beta$ decay experiments like SuperNEMO \cite{nemo}, KamLAND-Zen \cite{zen}, NEXT \cite{next1,next2},
	nEXO\cite{nexo} is, also, shown in Fig.(\ref{fig4}). $|M_{ee}|$ is found to be well within the sensitivity reach of these $0\nu\beta\beta$ experiments, in particular for inverted hierarchy. The model has imperative implication for inverted hierarchy, for example, the non-observation of this process at nEXO will rule out IH in this model. 
	
	The correlations of baryon asymmetry with the model parameters $|b|$, $|f|$ and $|d|$ are shown in Fig.(\ref{fig5}). It is evident from the figure that the predicted baryon asymmetry is in good agreement with the observed baryon asymmetry for NH whereas, for IH, there is accumulation of points away from the observed baryon asymmetry region. Infact IH is disallowed at 2.5$\sigma$ C.L.. Also, it is to be noted that right-handed neutrino mass scale is much below the GUT scale and will have negligible renormalization group evolution effects.
	
	\section{Conclusions}
	In conclusion, we have presented a model based on $A_4$ flavor symmetry augmented by $Z_3$ cyclic group within the framework of type-I and II seesaw mechanism in a minimal scenario of two right-handed neutrinos(2RHN). The type-II seesaw terms play a crucial role in generating non-degenerate neutrino masses and non-zero $\theta_{13}$(i.e. breaking $\mu-\tau$ symmetry). The Yukawa coupling involving scalar triplet $\Delta$ is responsible for breaking of $\mu-\tau$ symmetry and contributes to the baryogenesis. The resulted neutrino mass matrix retains magic symmetry and have non-degenerate mass eigenvalues. Using the data on neutrino mass and mixings, we have evaluated the model parameter space allowed by the constraints emanating from the magic symmetry within tolerance of $10^{-3}$. The effective Majorana neutrino mass $|M_{ee}|$ is well within the sensitivity reach of the $0\nu\beta\beta$ experiments. In particular, the non-observation of $0\nu\beta\beta$ decay at nEXO will rule out IH in this model.  We have, also, investigated the leptogenesis through approximated solutions of Boltzmann equations. We have worked in the approximation which is in accordance with the essential condition required for out-of-equilibrium decay of lightest right-handed neutrino. We find that the predicted baryon asymmetry is in good agreement with the observed baryon asymmetry for NH whereas IH is disallowed at 2.5$\sigma$ C.L..
	\section*{Acknowledgments}
M. K. acknowledges the financial support provided by the Department of Science and Technology(DST), Government of India vide Grant No. DST/INSPIRE Fellowship/2018/IF180327. The authors, also, acknowledge Department of Physics and Astronomical Science for providing necessary facility to carry out this work.


\begin{thebibliography}{99.}
		\bibitem{sno} Q. R. Ahmad \textit{et al.}, Phys. Rev. Lett. \textbf{87}, 071301 (2001).
		\bibitem{sno2} Q. R. Ahmad \textit{et al.}, Phys. Rev. Lett. 92,181301 (2004).
		\bibitem{kam}K. Eguchi \textit{et al.}, Phys. Rev. Lett. \textbf{90}, 021802 (2003).
		\bibitem{kam2}T. Araki et al., Phys. Rev. Lett. \textbf{94}, 081801 (2005).
		\bibitem{expt1}Y. Fukuda \textit{et al.}, Phys. Rev. Lett. \textbf{81}, 1562 (1998).
		\bibitem{expt1a}Y. Ashie \textit{et al.}, Phys. Rev. Lett. \textbf{93}, 101801 (2004).
		\bibitem{expt2}M. H. Ahn \textit{et al.}, Phys. Rev. Lett. \textbf{90}, 041801 (2003)
		\bibitem{expt2a} M. H. Ahn \textit{et al.}, Phys. Rev. Lett. \textbf{94}, 081802 (2005).
		\bibitem{expt3}G. D. Michael \textit{et al.}, Phys. Rev. D \textbf{97}, 191801 (2006).
		
		\bibitem{ino} S. Ahmed \textit{et al.}, Pramana J. Phys., \textbf{88}, 79 (2017).
		\bibitem{dune} R. Acciarri \textit{et al.}, arXiv:1512.06148[physics.ins-det].
		\bibitem{t2hk}  K. Abe \textit{et al.}, arXiv:1412.4673[physics.ins-det].
		\bibitem{nova} M. A. Acero \textit{et al.}, arXiv:1906.04907[hep-ex].
		\bibitem{katrin1} J. Angrik \textit{et al.}, KATRIN Design Report 2004, Wissenschaftliche Berichte, FZ Karlsruhe \textbf{7090}, (2004).
		\bibitem{katrin2}G. Drexlin V. Hannen, S. Mertens, and C. Weinheimer, Adv. High Energy Physics, \textbf{293986}, (2013).
		
		\bibitem{mm1} S. F. King, Rept. Prog. Phys. \textbf{67}, 107-158 (2004).
		\bibitem{mm2} Andre de Gouvea, Ann. Rev. Nucl. Part. Sci. \textbf{66}, 197-217 (2016). 
		\bibitem{mm3} S. F. King,  J. Phys. G: Nucl. Part. Phys. \textbf{42}, 123001 (2015). 
		
		\bibitem{tbm1} P. F. Harrison, D. H. Perkins and W. G. Scott, Phy. Lett. B \textbf{530}, 167 (2002).
		\bibitem{tbm2} P. F. Harrison and W. G. Scott, Phys. Lett. B \textbf{594}, 324 (2004).
		\bibitem{bm1} V. D. Barger, S. Pakvasa, T. J. Weiler and K. Whisnant, Phys. Lett. B \textbf{437}, 107 (1998). 
		\bibitem{bm2} G. Altarelli and F. Feruglio, J. High Energy Phys. \textbf{9811}, 021 (1998).
		\bibitem{gr1} Y. Kajiyama, M. Raidal, and A. Strumia, Phys. Rev. D \textbf{76}, 117301 (2007).
		\bibitem{gr2} W. Rodejohann, Phys. Lett. B \textbf{671}, 267 (2009).
		\bibitem{tm1} Carl H. Albright and Werner Rodejohann, Eur. Phys. J. C \textbf{62}, 599-608 (2009).
		\bibitem{tm2} Carl H. Albright, Alexander Dueck and Werner Rodejohann Eur. Phys. J. C \textbf{70}, 1099-1110 (2010).
		
		
		\bibitem{t2k} K. Abe \textit{et al.}, Phys. Rev. Lett. \textbf{107}, 041801 (2011).
		\bibitem{db}  F. P. An \textit{et al.}, Phys. Rev. Lett. \textbf{108}, 171803 (2012).
		\bibitem{dc} Y. Abe, \textit{et al.}, Phys. Rev. Lett., \textbf{108}, 131801 (2012).
		\bibitem{reno} J. Ahn, \textit{et al.}, Phys. Rev. Lett., \textbf{108}, 191802 (2012).
		
		
		
		\bibitem{bpr}Y. Akrami \textit{et al.}, arXiv:1807.06205[astro-ph.CO].
		
		
		\bibitem{sakharov} A. Sakharov, Pisma Zh. Eksp. Teor. Fiz. \textbf{5}, 32 (1967).
		
		

		
		\bibitem{magic}K. S. Channey and S. Kumar,  J. Phys. G \textbf{46},  015001 (2019). 
		\bibitem{hybrid}S. Dev, R.R. Gautam, L. Singh, Phys. Rev. D \textbf{87}, 073011 (2013).
		\bibitem{data} P. F. de Salas \textit{et al.}, Phys. Lett. B \textbf{782}, 633-640 (2018).
		
		\bibitem{eps1} S. Antusch and S. F. King, Phys. Lett. B \textbf{597}, 199 (2004).
		\bibitem{eps3} T. Hambye and G. Senjanovic, Phys. Lett. B \textbf{582}, 73 (2004).
		\bibitem{buch} W. Buchmuller, P. Di Bari and M. Plumacher, Annals Phys. \textbf{315}, 305-351 (2005).
		\bibitem{cosmo} S Vagnozzi \textit{et al.}, Phys. Rev. D \textbf{96}, 123503 (2017).
		\bibitem{dibound}S. Davidson and A. Ibarra, Phys. Lett. B \textbf{535}, 25 (2002).
		\bibitem{nemo} A. S. Barabash, J. Phys. Conf. Ser. \textbf{375}, 042012 (2012).
		\bibitem{zen} A. Gando \textit{et al.}, Phys. Rev. Lett. \textbf{117}, 082503 (2016).
		\bibitem{next1} F. Granena \textit{et al.}, arXiv:0907.4054[hep-ex].
		\bibitem{next2} J. J. Gomez-Cadenas \textit{et al.}, Adv. High Energy Phys. \textbf{2014}, 907067 (2014).
		\bibitem{nexo} C. Licciardi, J. Phys. Conf. Ser. \textbf{888}, 012237 (2017).
		
	\end{thebibliography}
\end{document}